\documentclass{article}

     \usepackage[nonatbib,preprint]{neurips_2021}

\usepackage[utf8]{inputenc} 
\usepackage[T1]{fontenc}    
\usepackage{hyperref}       
\usepackage{url}            
\usepackage{booktabs}       
\usepackage{amsfonts}       
\usepackage{nicefrac}       
\usepackage{microtype}      
\usepackage[noend,linesnumbered,ruled,vlined]{algorithm2e}
\usepackage{xcolor}
\usepackage{xspace}
\usepackage{amsthm}
\usepackage{amsmath}
\usepackage{enumitem}
\usepackage{pgfplots}
\usepackage{subcaption}
\usepackage{wrapfig}
\usepackage{systeme}
\usepackage{mathtools}

\usepgfplotslibrary{fillbetween}

\newcommand{\fedchi}{\textsc{Fed}-$\chi^2$\xspace}
\DeclareMathOperator{\sgn}{sgn}
\newtheorem{theorem}{Theorem}
\newtheorem{definition}{Definition}
\newtheorem{lemma}{Lemma}

\newcommand{\F}{Fig.}
\renewcommand{\S}{Sec.}
\newcommand{\A}{Alg.}

\title{\fedchi: Privacy Preserving Federated Correlation Test}

\author{%
  Lun Wang\thanks{The authors contribute equally to this paper.} \\
  University of California, Berkeley\\
  \texttt{wanglun@berkeley.edu} \\
  \And Qi Pang\footnotemark[1]\\
  Hong Kong University of Science and Technology\\
  \texttt{qpangaa@cse.ust.hk}
  \AND Shuai Wang\\
  Hong Kong University of Science and Technology\\
  \texttt{shuaiw@cse.ust.hk}
  \And Dawn Song\\
  University of California, Berkeley\\
  \texttt{dawnsong@cs.berkeley.edu}
}

\begin{document}

\maketitle

\begin{abstract}
In this paper, we propose the first secure federated $\chi^2$-test protocol \fedchi.
To minimize both the privacy leakage and the communication cost, we recast
$\chi^2$-test to the second moment estimation problem and thus can take advantage of stable projection to encode the local information in a short vector.
As such encodings can be aggregated with only summation, secure aggregation can be naturally applied to hide the individual updates.
We formally prove the security guarantee of \fedchi that the joint distribution is hidden in a subspace with exponential possible distributions. 
%
Our evaluation results show that \fedchi achieves negligible accuracy drops with small client-side computation overhead. 
In several real-world case studies, the performance of \fedchi is comparable to the centralized $\chi^2$-test.
%
\end{abstract}

\vspace{-0.2em}
\section{Introduction}
\label{sec:intro}
\vspace{-0.6em}

Correlation test, as its name indicates, refers to the process of using observational data to test the correlation between two random variables.
It serves as a basic building block in many real-world applications such as feature selection~\cite{zheng2004feature}, cryptanalysis~\cite{nyberg2001correlation}, causal graph discovery~\cite{spirtes2000constructing}, empirical finance~\cite{ledoit2008robust,kim2015significance}, medical studies~\cite{kassirer1983teaching} and genomics~\cite{wilson1999rapid,dudoit2003multiple}.
The observational data used for correlation test might contain sensitive information such as genome information so it is risky to collect the data in a centralized manner.
To address, we turn to the federated setting where each client holds its data locally and communicates with a centralized server to calculate a function. 
Note that the communication should contain as little information as possible.
Otherwise, the server might be able to infer sensitive information from the communication.

In this paper, we study a representative correlation test method, $\chi^2$-test,
in the federated learning setting. Specifically, we aim to minimize privacy
leakage with modest communication cost. To conduct $\chi^2$-test trivially in a
collaborative setting, the clients upload their raw data in plaintext to the
centralized server, which is communication-efficient but completely reveals the
clients' private information. On the other end of the spectrum, the clients
might run secure multiparty computation (MPC) under the coordination of a
server. This way, clients can joint launch $\chi^2$-test with their local data
without revealing the data to the server. However, general-purpose MPC incurs
considerable computation and communication overhead, which are usually
intolerable in real-world federated settings, e.g., running federated clients on
mobile devices.

To address the dilemma, this research designs a new federated protocol
specifically optimized for the application of correlation test; our novel
federated protocol is \emph{computation/communication-efficient} and \emph{with
  minor privacy leakage}.
We first show that $\chi^2$-test can be recast to the problem of frequency moments estimation.
To approximate frequency moments in a federated manner, each client encodes its raw data into a low-dimensional vector.
The server then aggregates the encodings and decodes it to approximate the frequency moments.
We choose stable random projection~\cite{indyk2006stable,vempala2005random,li2008estimators} as the encoding scheme, given it only requires
summation to aggregate and is thus friendly to secure aggregation
techniques~\cite{bonawitz2017practical,bell2020secure}.
Secure aggregation hides each client's update in the final aggregate cryptographically and hence provides rigorous security guarantees.

The information leakage of \fedchi\ is much less than the joint distribution because the global contingency table of $\chi^2$-test is hidden in a subspace of exponential size and each client's raw data is perfectly hidden with secure aggregation~\cite{bonawitz2017practical,bell2020secure}.
Furthermore, the communication cost is much lower than general-purpose MPC.
%
In particular, the communication cost of \fedchi\ is $\mathcal{O}(\log n + m_x +
m_y + \ell)$ per client,
%
%
where $n$ is the number of clients, $m_x$ and $m_y$ are the dimensions of the
contingency table, $\ell$ is the encoding size, and $m_xm_y\gg\ell,
m_xm_y\gg \log n$.
To compare, if we use the delegation model where the clients secretly share
their inputs with two servers and let them run the MPC protocol, then the
communication cost is at least $\mathcal{O}(m_xm_y)$ for uploading the complete
contingency tables.
%

Our evaluation on four synthetic datasets and 16 real-world datasets shows that
\fedchi\ can substitute centralized $\chi^2$-test with small multiplicative
error and computation overhead.
Furthermore, we evaluate the performance of
\fedchi\ on three real-world cases: feature selection, cryptanalysis, and online
false discovery rate control. The results show that \fedchi\ can achieve
comparable performance to centralized $\chi^2$-test on corresponding
task-specific metrics. In summary, we make the following contributions:

\vspace{-0.5em}
\noindent {\tikz\draw[black,fill=black] (-0.5em,-0.5em) circle (-0.15em);} We
propose \fedchi, the first secure federated $\chi^2$-test protocol. \fedchi is
computation- and communication-efficient, and leaks much less information than trivially deploying secure aggregation.

\vspace{-0.5em}
 
\noindent {\tikz\draw[black,fill=black] (-0.5em,-0.5em) circle (-0.15em);} 
\fedchi recasts $\chi^2$-test to frequency moments estimation and incorperates stable random projection and secure aggregation techniques to minimize communication cost and privacy leakage.
We give formal security proof and utility analysis of \fedchi.

\vspace{-0.5em}
\noindent {\tikz\draw[black,fill=black] (-0.5em,-0.5em) circle (-0.15em);} We
evaluate \fedchi both analytically and in real use cases. The results show that
\fedchi can substitute centralized $\chi^2$-test with comparable accuracy.

\vspace{-0.8em}
\section{Related Work}
\label{sec:background}
\vspace{-0.6em}
Bonawitz et al.~\cite{bonawitz2017practical} proposed the well-celebrated secure
aggregation protocol to calculate linear functions in the federated setting at
modest cost. 
Since then, it has seen many variants and improvements~\cite{truex2019hybrid,xu2019hybridalpha,so2021turbo,bell2020secure,choi2020communication}.
For instance, Truex et al.~\cite{truex2019hybrid} and Xu et al.~\cite{xu2019hybridalpha}
employed advanced crypto tools such as threshold homomorphic encryption and
functional encryption for secure aggregation. 
So et al.~\cite{so2021turbo}
proposed {\sc TurboAGG} which combines secure sharing with erasure codes for
better dropout tolerance. 
Bell et al.~\cite{bell2020secure} and Choi et
al.~\cite{choi2020communication} replaced the complete graph in secure
aggregation with either a sparse random graph or a low-degree graph to enhance communication efficiency.

Secure aggregation is deployed in various applications. Agarwal et
al.~\cite{agarwal2018cpsgd} added binomial noise to local gradients, achieving
both differential privacy and communication efficiency. Wang et
al.~\cite{wang2020d2p} replaced the binomial noise with discrete Gaussian noise,
which is shown to exhibit better composability. Kairouz et
al.~\cite{kairouz2021distributed} proved that the sum of discrete Gaussian is
close to discrete Gaussian, thus discarding the common random seed assumption
from~\cite{wang2020d2p}.
The above three works all incorporate secure
aggregation in their protocols to reduce the noise scale needed for differential
privacy.
Chen et al.~\cite{chen2020training} added an extra public parameter to
each client to force them to train in the same way, thus malicious clients can be
detected during aggregation. 

Private federated statistics have also drawn numerous attention in the past few years. 
Bassily et al.~\cite{bassily2015local} studied the problem of differentially private frequency estimation in the local model. 
Chen et al.~\cite{chen2020breaking} proposed new schemes for federated private mean estimation and heavy hitter estimation. 
Zhu et al.~\cite{zhu2020federated}
proposed distributed and privacy-preserving algorithms for heavy hitter estimation. 
Acharya et al.~\cite{acharya2021estimating} designed a federated protocol to estimate sparse discrete distributions. 
%
%
%
Blocki et al.~\cite{blocki2012johnson} studied 2nd-stable projection under the name of Johnson-Lindenstrauss transform and proved it is differentially private.
%
%
However, these results are based on the confidentiality of the projection matrix and thus do not apply to our algorithm in which the matrix is public.

\vspace{-0.8em}
\section{Federated Correlation Test with Minimal Leakage}
\label{sec:fedchi-proof}
\vspace{-0.6em}

In this section, we elaborate on the design of \fedchi, a privacy-preserving
federated protocol for $\chi^2$-test. \S~\ref{subsec:threat-model} first
formalizes the problem, establishes the notation system and introduces the threat
model.
In \S~\ref{subsec:recast}, we recast $\chi^2$-test to a frequency moments
estimation under the federated setting, and consequently, we are able to
leverage stable projection to encode each client's local information
(\S~\ref{subsec:encoding}), and further aggregate them with secure aggregation
(\S~\ref{subsec:protocol}). \S~\ref{subsec:security} presents security proof,
utility analysis, communication analysis, and computation analysis of \fedchi.

\vspace{-0.8em}
\subsection{Problem Setup}
\label{subsec:threat-model}
\vspace{-0.6em}

We now formulate the problem of federated correlation test and establish the
notation system. We use $[n]$ to denote $\{1, \cdots, n\}$. We denote vectors
with bold lower-case letters (\emph{e.g.}, $\textbf{a}, \textbf{b}, \textbf{c}$)
and matrices with bold upper-case letters (\emph{e.g.}, $\textbf{A}, \textbf{B},
\textbf{C}$).

We consider a population of $n$ clients $\mathcal{C}=\{c_i\}_{i\in[n]}$. Each
client has one share of local data composed of the triplets $\mathcal{D}_i=\{(x,
y, v^{(i)}_{xy})\}, x\in\mathcal{X}, y\in\mathcal{Y}, v^{(i)}_{xy}\in\{-M,
\cdots, M\}$, where $|\mathcal{X}|=m_x$ and $|\mathcal{Y}|=m_y$ are finite
domains, $M$ is the maximum value $|v^{(i)}_{xy}|$ can be.
The global dataset is defined as $\mathcal{D}=\{(x, y,
v_{xy}):v_{xy}=\sum_{i\in[n]}v^{(i)}_{xy}\}$.
For the ease of presentation, we define the marginal distributions $v_x=\sum_{y\in[|\mathcal{Y}|]}v_{xy},
v_y=\sum_{x\in[|\mathcal{X}|]}v_{xy}$, and $v=\sum_{x\in[|\mathcal{X}|],
  y\in[|\mathcal{Y}|]}v_{xy}$.
Besides, we define $\bar{v}_{xy}=\frac{v_x\times v_y}{v}$, denoting the
expectation of $v_{xy}$ if $x$ and $y$ are uncorrelated.
We define $m=m_xm_y$ and use an indexing function $\mathbb{I}:
[m_x]\times[m_y]\rightarrow [m]$ to obtain a uniform indexing given the indexing
of each variable. Thus, a centralized server $\mathcal{S}$ calculates the
statistics for $\chi^2$-test $s_{\chi^2}(\mathcal{D})=\sum_{x\in[|\mathcal{X}|],y\in[|\mathcal{Y}|]}\frac{(v_{xy}-\bar{v}_{xy})^2}{\bar{v}_{xy}}$ on the global dataset to decide whether $\mathcal{X}$ and
$\mathcal{Y}$ are correlated without collecting the raw data from clients.
%

%

Overall, conducting correlation test losslessly in the federated setting is
highly costly and impractical when using previous approaches like
MPC~\cite{boyle2015large,damgaard2012multiparty}. Hence, in this research, we
trade accuracy for efficiency, as long as the estimation error is small with
high probability. Formally, if \fedchi\ outputs $\hat{s}_{\chi^2}$, whose
corresponding $\chi^2$-test output is $s_{\chi^2}$ in the standard centralized
setting, the following constraint should be satisfied with small $\epsilon$ and
$\delta$.
\vspace{-0.2em}
\begin{equation*}
\mathbb{P}[(1-\epsilon)s_{\chi^2}\leq\hat{s}_{\chi^2}\leq(1+\epsilon)s_{\chi^2}]\geq 1 - \delta 
\end{equation*}


\vspace{-1em}
\paragraph{Threat Model.}
We assume that the centralized server $\mathcal{S}$ is honest but curious. That
is, it honestly follows the protocol due to regulation or reputational
pressure but is curious to learn additional private information from the
legitimate updates made by the clients for profit or surveillance.
Hence, the updates from the clients should leak as little private information as
possible.

On the other hand, we assume totally honest clients. Specifically, we do not consider client-side adversarial attacks (e.g., data poisoning attacks~\cite{bagdasaryan2020backdoor,bhagoji2019analyzing}),
and we do not allow the server to corrupt any client. We also do not consider the
dropout of clients during the execution.

\vspace{-0.8em}
\subsection{From Correlation Test to Frequency Moments Estimation}
\label{subsec:recast}
\vspace{-0.6em}

We now recast correlation test to frequency moments estimation. Frequency moments
estimation refers to the following problem. Given a set of key-value pairs
$\mathcal{S}=\{k_i, v_i\}_{i\in[n]}$, we re-organize it into a histogram
$\mathcal{H}=\{k_j, v_j=\sum_{k_i=k_j, i\in[n]}v_i\}$, and estimate
the $\alpha^{th}$ frequency moments as $F_\alpha=\sum_j v_j^\alpha$.


$\chi^2$-test can thus be recast to a $2^{nd}$ frequency moments estimation
problem as follows.
\vspace{-0.2em}
\begin{equation*}
s_{\chi^2}(\mathcal{D})=\sum_{x,y}\frac{(v_{xy}-\bar{v}_{xy})^2}{\bar{v}_{xy}}=\sum_{x, y}(\frac{v_{xy}-\bar{v}_{xy}}{\sqrt{\bar{v}_{xy}}})^2
\end{equation*}

\vspace{-0.2em}

In federated setting, each client $c_i$ holds a local dataset $\mathcal{D}_i=\{(x, y, v_{xy}^{(i)})\}$ and computes a vector $\textbf{u}_i$, where $\textbf{u}_i [\mathbb{I}(x,y)] = \frac{v_{xy}^{(i)}-\bar{v}_{xy}/n}{\sqrt{\bar{v}_{xy}}}$ and $\textbf{u}_i$ has $m$ elements. 
Thus, the challenge in federated $\chi^2$-test becomes to calculate the
following without leaking the private information $\textbf{u}_i$ and $\sum_{i
  \in [n]}\textbf{u}_i$.

\vspace{-0.2em}
\begin{equation*}
s_{\chi^2}(\mathcal{D}) = \sum_{x,y}(\frac{v_{xy}-\bar{v}_{xy}}{\sqrt{\bar{v}_{xy}}})^2 = ||\sum_{i \in [n]}\textbf{u}_i||_2^2
\end{equation*}

\vspace{-0.2em}

%



\vspace{-0.8em}
\subsection{Encoding with Stable Projection \& Decoding with Geometric Mean Estimator}
\label{subsec:encoding}
\vspace{-0.6em}
To address the above challenges and integrate the algorithm into secure aggregation protocols naturally, we use \textit{stable random projection}~\cite{indyk2006stable,vempala2005random} and \textit{geometric mean estimator}~\cite{li2008estimators} to efficiently compute the $l_2$ norm of the data. 
We first introduce stable distributions and then present the encoding and decoding algorithms.

\begin{definition}[$\alpha$-stable distribution]
A random variable $X$ follows an $\alpha$-stable distribution $\mathcal{Q}_{\alpha, \beta, F}$ if its characteristic function is as follows.
\vspace{-0.2em}
\begin{equation*}
\phi_X(t) = \exp(-F|t|^p(1-\sqrt{-1}\beta\sgn(t)\tan(\frac{\pi\alpha}{2})))
\end{equation*}
\vspace{-0.2em}
, where $F$ is the scale to the $\alpha^{th}$ power and $\beta$ is the skewness.

\label{def:stable-dist}
\end{definition}

\vspace{-0.3em}
$\alpha$-stable distribution is named due to its property called $\alpha$-stability.
Briefly, the sum of independent $\alpha$-stable variables still follows an
$\alpha$-stable distribution with a different scale.

\begin{definition}[$\alpha$-stability]
If random variables $X\sim\mathcal{Q}_{\alpha, \beta, 1}, Y\sim\mathcal{Q}_{\alpha, \beta, 1}$ and $X$ and $Y$ are independent, then $C_1X+C_2Y\sim\mathcal{Q}_{\alpha,\beta, C_1^\alpha+C_2^\alpha}$.
\label{def:stability}
\end{definition}

\vspace{-0.3em}
We borrow the genius idea from Indyk's well-celebrated paper~\cite{indyk2006stable} to encode the frequency moments in the scale parameter of a stable distribution defined in Definition~\ref{def:stable-dist}.
%
To encode the local dataset $\mathcal{D}_i=\{(x, y, v_{xy}^{(i)})\}$, each client $c_i$, $i \in [n]$, is given by the server an $\ell \times m$ projection matrix $\textbf{P}$ whose values are drawn independently from an $\alpha$-stable distribution $\mathcal{Q}_{\alpha, \beta, 1}$, where $\ell$ is the encoding size and $m = m_xm_y$.
The client re-organizes the data into a vector $\textbf{u}_i$, where
$\textbf{u}_i[\mathbb{I}(x, y)]=v_{xy}^{(i)}$. Then, the client projects
$\textbf{u}_i$ to $\textbf{e}_i=\textbf{P}\times\textbf{u}_i$ as the encoding (line 2 in \A~\ref{alg:encode}).

To decode, the server first sums the encodings up $\textbf{e} = \sum_{i \in [n]} \textbf{e}_i$ and estimates the scale of the
variables in the aggregated encoding with an unbiased geometric mean
estimator~\cite{li2008estimators} in line 5 of \A~\ref{alg:encode}.
%
%
According to the $\alpha$-stability defined in Definition~\ref{def:stability}, 
every element $e_k$ in $\textbf{e}$, $k \in [\ell]$, follows this stable distribution: $e_k \sim \mathcal{Q}_{2, 0, ||\sum_{i \in [n]}\textbf{u}_i||_2^2}$. 
Thus, the $l_2$ norm can be estimated by calculating the scale of the distribution $\mathcal{Q}_{2, 0, ||\sum_{i \in [n]}\textbf{u}_i||_2^2}$ with $\textbf{e}$ containing $\ell$ elements. 

During above process, $\textbf{u}_i$ and $\sum_{i \in [n]} \textbf{u}_i$ are not
leaked and the aggregation among clients is calculated with only summation.
Thus, secure aggregation protocols can be naturally applied, as will be
discussed in \S~\ref{subsec:security}.
%
Furthermore, $\ell = \frac{c}{\epsilon^2}\log (1/\delta)$ suffices to guarantee that any $l_\alpha$ distance can be approximated with a $1 \pm \epsilon$ factor and a probability no less than $1 - \delta$.
We will further analyze the utility of \fedchi in \S~\ref{subsec:utility}.

\begin{figure}
\vspace{-1.5em}
\begin{algorithm}[H]
  \footnotesize
  \caption{The encoding and decoding scheme for federated frequency moments estimation. Note that the encoding and decoding themselves do not provide any security guarantee.}
  \label{alg:encode}
  \DontPrintSemicolon
  \SetKwFunction{FEncode}{\textsc{Encode}}
  \SetKwFunction{FDecode}{\textsc{Decode}}
  \SetKwFunction{FG}{\textsc{GeometricMeanEstimator}}
  \SetKwProg{Fn}{Function}{:}{}
  \Fn{\FEncode{$\textbf{P}$, $\textbf{u}_i$}}{
        \KwRet $\textbf{P}\times \textbf{u}_i$\;
  }
  \Fn{\FG{$\textbf{e}$}}{
        $\triangleright$ $\ell$: Encoding size.\;
        $\hat{d}_{(2), gm} \leftarrow \frac{\prod_{k=1}^\ell|\textbf{e}_k|^{2/\ell}}{(\frac{2}{\pi}\Gamma(\frac{2}{\ell}) \Gamma(1 - \frac{1}{\ell}) \sin(\frac{\pi}{\ell}))^\ell}$ \;
        \KwRet $\hat{d}_{(2), gm}$
  }
  \Fn{\FDecode{$\textbf{e}$}}{
        \KwRet \FG{$\textbf{e}$}\;
  }
\end{algorithm}
\vspace{-2em}
\end{figure}
%

\vspace{-0.8em}
\subsection{Secure Federated Correlation Test}
\label{subsec:protocol}
\vspace{-0.6em}

The complete protocol for \fedchi\ is presented in \A~\ref{alg:main}. 
Firstly, the marginal distributions $v_x, v_y$ and $v$ are calculated with
secure aggregation and broadcasted to all clients (lines 1--6 in
\A~\ref{alg:main}). And this step can be omitted if the marginal distribution is
already known.
Then, on the server side, an $\ell \times m$ projection matrix $\textbf{P}$ is
sampled from $\alpha$-stable distributions $\mathcal{Q}_{2, 0, 1}^{\ell\times
  m}$.
The projection matrix is broadcasted to all clients (lines 8--10 in
\A~\ref{alg:main}).
For each client $c_i$, the local data is re-organized into $\textbf{u}_i$ and
projected to $\textbf{e}_i$ as encoding (lines 11--14 in \A~\ref{alg:main}).
Then, the encoding results from clients are aggregated with secure aggregation
(line 15 in \A~\ref{alg:main}).
Finally, the server gets the $\chi^2$-test result by the decoding algorithm as
described in \A~\ref{alg:encode} (line 17 in \A~\ref{alg:main}).

\begin{figure}
\vspace{-1em}
\begin{algorithm}[H]
    \footnotesize
    \DontPrintSemicolon
    \SetKwBlock{Round}{Round}{}
    \SetKwBlock{Server}{Server}{}
    \SetKwBlock{Client}{Client}{}
    \SetKwFunction{FEncode}{\textsc{Encode}}
    \SetKwFunction{FDecode}{\textsc{Decode}}
    \SetKwFunction{FInit}{\textsc{InitSecureAgg}}
    \SetKwFunction{FSecureagg}{\textsc{SecureAgg}}
    \SetKw{Return}{return}
    \SetAlgoLined
    \Round(1: Reveal the marginal distribution){
        \FInit{n} \tcp*{$n$ is the client number.}
        \lFor{$x\in[m_x]$}{$v_{x} = $ \FSecureagg{$\{v_{x}^{(i)}\}_{i\in[n]}$} }
        \lFor{$y\in[m_y]$}{$v_{y} = $ \FSecureagg{$\{v_{y}^{(i)}\}_{i\in[n]}$} }
        \Server{
            Calculate $v=\sum_x{v_x}$ and broadcast $v$, $\{v_{x}\}$ and $\{v_{y}\}$ to all the clients.
        }
    }
    \Round(2: Approximate the statistics){
        \Server{
            {Sample the projection matrix $\textbf{P}$ from $\mathcal{Q}_{2, 0, 1}^{\ell\times m}$}\;
            Broadcast the projection matrices to the clients\;
        }
        \Client($c_i, i \in [n]$){
            Calculate $\bar{v}_{xy}=\frac{v_{x} v_{y}}{v}$\;
            {Prepare $\textbf{u}_i$ s.t. $\textbf{u}_i[\mathbb{I}(x,y)]=\frac{v_{xy}^{(i)}-\bar{v}_{xy}/n}{\sqrt{\bar{v}_{xy}}}$\; 
            Calculate $\textbf{e}_i=$ \FEncode{$\textbf{P}, \textbf{u}_i$} }\;
        }
        {$\textbf{e}=$ \FSecureagg{$\{\textbf{e}_i\}_{i\in[n]}$} }\;
        \Server{
            {$\hat{s}_{\chi^2}=$ \FDecode{$\textbf{e}$} }\;
        }
    }
\caption{\fedchi: secure federated $\chi^2$-test. $\textsc{SecureAgg}$ is a remote procedure that receives inputs from the clients and returns the summation to the server. $\textsc{InitSecureAgg}$ is the corresponding setup protocol deciding the communication graph and other hyper-parameters.}
\label{alg:main}
\end{algorithm}
\vspace{-2em}
\end{figure}

%

%
\vspace{-0.8em}
\subsection{Security Analysis}
\label{subsec:security}
\vspace{-0.6em}

As we have reviewed in \S~\ref{sec:background}, the secure aggregation protocol
is well studied and many well-celebrated secure aggregation protocols have been
proposed~\cite{bonawitz2017practical,truex2019hybrid,xu2019hybridalpha,so2021turbo,bell2020secure,choi2020communication}.
In this paper, we choose to use a state-of-the-art secure aggregation protocol
presented by Bell et al.~\cite{bell2020secure}, which replaces the complete
graph with a sparse random graph to enhance communication efficiency.
We emphasize that \fedchi\ can incorporate other popular secure aggregation
protocols; we use the protocol proposed in~\cite{bell2020secure} for its high
communication efficiency and low computation cost. We now prove the security
enforced by \A~\ref{alg:main} using Theorem~\ref{thm:sec}.
We then clarify the security implication of Theorem 1 --- indistinguishability
--- using standard simulation proof~\cite{Lindell2017}. We further discuss the
marginal privacy leakage implied by \A~\ref{alg:main}.

\begin{theorem}[Security]
Let $\Pi$ be an instantiation of \A~\ref{alg:main} with the secure aggregation protocol in \A~\ref{alg:secureagg} of Appendix~\ref{sec:secureagg} with 
cryprographic security parameter $\lambda$.
There exists a PPT simulator \textsc{Sim} such that for all clients $\mathcal{C}$, the number of clients $n$, all the marginal distributions $\{v_x\}, \{v_y\}$, and all the encodings $\{\textbf{e}_i\}$ ($\{(\textbf{e}_i^+\}, \{\textbf{e}_i^-\}, \{\textbf{w}_i)\}$), the output of \textsc{Sim} is indistinguishable from the view of the real server $\Pi_\mathcal{C}$ in that execution, \emph{i.e.},  $\Pi_\mathcal{C}\approx_{\lambda}\textsc{Sim}(\sum\textbf{e}_i, n)$ ($\Pi_\mathcal{C}\approx_{\lambda}\textsc{Sim}(\sum\textbf{e}_i^+, \sum\textbf{e}_i^-, \sum\textbf{w}_i, n, \{v_x\}, \{v_y\})$).
\label{thm:sec}
\end{theorem}

\vspace{-1em}
\begin{proof}[Proof for Theorem~\ref{thm:sec}]

To prove Theorem~\ref{thm:sec}, we need the following lemma.
\begin{lemma}[Security of secure aggregation protocol]
Let \textsc{SecureAgg} be the secure aggregation protocol in \A~\ref{alg:secureagg} of Appendix~\ref{sec:secureagg} instantiated with cryprographic security parameter $\lambda$. 
There exists a PPT simulator \textsc{SimSA} such that for all clients $\mathcal{C}$, the number of clients $n$, and all inputs $\mathcal{X}=\{\textbf{e}_i\}_{i\in[n]}$, the output of \textsc{SimSA} is perfectly indistinguishable from the view of the real server, \emph{i.e.}, $\textsc{SecureAgg}_\mathcal{C}\approx_{\lambda}\textsc{SimSA}(\sum_{i\in[n]} \textbf{e}_i, n)$.

\label{lem:secagg}

\end{lemma}
\vspace{-1em}
Lemma~\ref{lem:secagg} comes from the security analysis for the secure aggregation protocol (Theorem 3.6 in~\cite{bell2020secure}) which illustrates that the secure aggregation protocol securely hides the individual information in the aggregated result. 
With this lemma, we are able to prove the theorem for federated $\chi^2$-test by presenting a sequence of hybrids which start from the real protocol execution and transition to the simulated execution. 
We prove that every two consecutive hybrids are indistinguishable so all the hybrids are indistinguishable according to transitivity.
%
\vspace{-1em}
\begin{itemize}[leftmargin=15mm]
    \setlength\itemsep{0em}
    \item[\textsc{Hyb}$_1$] This is the view of the server in the real protocol execution, $\textsc{Real}_\mathcal{C}$.
    \vspace{-0.5em}
    \item[\textsc{Hyb}$_2$] In this hybrid, we replace the view during the execution of each $\textsc{SecureAgg}(\{v_{x}^{(i)}\}_{i\in[n]})$ in line 3 of \A~\ref{alg:main} with the output of $\textsc{SimSA}(v_x, n)$ one by one. According to Lemma~\ref{lem:secagg}, each replacement does not change the indistinguishability. Hence, \textsc{Hyb}$_2$ is indistinguishable from \textsc{Hyb}$_1$.
    \vspace{-0.5em}
    \item[\textsc{Hyb}$_3$] Similar to \textsc{Hyb}$_2$, we replace the view during the execution of each $\textsc{SecureAgg}(\{v_{y}^{(i)}\}_{i\in[n]})$ in line 4 of \A~\ref{alg:main} with the output of $\textsc{SimSA}(v_y, n)$ one by one. According to Lemma~\ref{lem:secagg}, \textsc{Hyb}$_3$ is indistinguishable from \textsc{Hyb}$_2$. 
    \vspace{-0.5em}
    \item[\textsc{Hyb}$_4$] In this hybrid, we replace the view during the execution of $\textsc{SecureAgg}(\{\textbf{e}_i\}_{i\in[n]})$ in line 15 of \A~\ref{alg:main} with the $\textsc{SimSA}(\sum\textbf{e}_i, n)$. This hybrid is the output of $\textsc{Sim}$. According to Lemma~\ref{lem:secagg}, \textsc{Hyb}$_4$ is indistinguishable from \textsc{Hyb}$_3$. \qedhere
\end{itemize}
\end{proof}

\vspace{-0.8em}
\paragraph{Remark: what does \A~\ref{alg:main} leak?}
By Theorem~\ref{thm:sec}, we show that \fedchi\ leaks no more than a linear equation system given by
%
%
%
\begin{equation*}
\left\{
\begin{array}{cr}
    \textbf{P} \times \textbf{v} &= \textbf{e}^T \\[0.2em]
    \textbf{J}_{1,m_y} \times \textbf{V}^T &= \textbf{v}_x^T \\[0.2em]
    \textbf{J}_{1,m_x} \times \textbf{V} &= \textbf{v}_y^T
\end{array}
\right.
\end{equation*}
, where $\textbf{J}_{1,m_x}$ and $\textbf{J}_{1,m_y}$ are $1 \times m_x$ and $1 \times m_y$ unit matrices, $\textbf{V}$ is an $m_x \times m_y$ matrix whose elements are $\{v_{xy}\}$, and $\textbf{v}$ is a vector flattened by $\textbf{V}$.
To understand this, information leaked by \fedchi includes the estimation $\textbf{e}$ and marginal distributions $\textbf{v}_x$ and $\textbf{v}_y$. 
In the following theorem, we show an important fact that the above equation system has a solution space of exponential size and thus the real joint distribution is securely hidden in the space. 
%
We deem that \A~\ref{alg:main} practically enforces privacy given the considerably large size of the solution space.

\begin{theorem}
Given a projection matrix $\textbf{P}\in\mathbb{Z}_q^{\ell\times m}$, $\{v_x\}$, $\{v_y\}$ and $\textbf{e}$, there are at least $q^{m-\ell-m_x-m_y}$ feasible choices of $\{v_{xy}\}$.
\label{thm:space}
\end{theorem}

\vspace{-1em}
\begin{proof}[Proof sketch for Theorem~\ref{thm:space}]
The given information forms a linear equation with $m_x + m_y + \ell$ equations as shown above.
Given $m>m_x+m_y+\ell$, the rank of the coefficient matrix is at most $m_x+m_y+\ell$.
Solving the equation with Smith normal form, we know that the solution space is at least $(m-m_x-m_y-\ell)$-dimensional.
With the following standard lemma, we manage to prove Theorem~\ref{thm:space}. 

\begin{lemma}
There are $q^{r\times c}$ vectors in the subspace of $\mathbb{Z}_q^{r\times c}$. \qedhere
\end{lemma}
\end{proof}


\vspace{-1em}
\subsection{Utility Analysis}
\label{subsec:utility}
\vspace{-0.6em}

In this section, we conduct the utility analysis in terms of multiplicative
error. We show that the output of \fedchi, $\hat{s}_{\chi^2}$, is a fairly
accurate approximation (parameterized by $\epsilon$) to the correlation test
output $s_{\chi^2}$ in the standard centralized setting with high probability
parameterized by $\delta$.

\begin{theorem}[Utility]
Let $\Pi$ be an instantiation of \A~\ref{alg:main} with secure
aggregation protocol in \A~\ref{alg:secureagg} of Appendix~\ref{sec:secureagg}. $\Pi$ is parameterized with
$\ell=\frac{c}{\epsilon^2}\log(1/\delta)$ for
some constant $c$. After
executing $\Pi_\mathcal{C}$ on all clients $\mathcal{C}$, the server yields $\hat{s}_{\chi^2}$, whose
distance to the accurate correlation test output $s_{\chi^2}$ is bounded with
high probability as follows:
\begin{equation*}
\mathbb{P}[\hat{s}_{\chi^2}<(1-\epsilon)s_{\chi^2} \vee \hat{s}_{\chi^2}>(1+\epsilon)s_{\chi^2}]\leq \delta
\end{equation*}

\label{thm:utility}

\end{theorem}

\vspace{-1em}
\begin{proof}[Proof sketch for Theorem~\ref{thm:utility}]
The proof is generally straightforward given the following lemma from~\cite{li2008estimators}.


\begin{lemma}[Tail bounds of geometric mean estimator~\cite{li2008estimators}]
The right tail bound of geometric mean estimator is:
\begin{equation*}
\mathbb{P}(\hat{s}_{\chi^2} - s_{\chi^2} > \epsilon s_{\chi^2}) \leq \exp(-\ell \frac{\epsilon^2}{G_{R, \alpha, \epsilon}})
\end{equation*}
, where $\frac{\epsilon^2}{G_{R, \alpha, \epsilon}} = C_1 \log (1 + \epsilon) - C_1 \gamma_e(\alpha - 1) - \log (\frac{2}{\pi} \Gamma(\alpha C_1) \Gamma(1 - C_1) \sin (\frac{\pi \alpha C_1}{2}))$, $\alpha = 2$ in our setting, $C_1 = \frac{2}{\pi} \tan^{-1}(\frac{\log (1 + \epsilon)}{(2 + \alpha^2)\pi/6})$, and $\gamma_e = 0.577215665...$ is the Euler's constant.

The left tail bound of the geometric mean estimator is:
\begin{equation*}
\mathbb{P}(\hat{s}_{\chi^2} - s_{\chi^2} < -\epsilon s_{\chi^2}) \leq \exp (-\ell\frac{\epsilon^2}{G_{L, \alpha, \epsilon, \ell_0}})
\end{equation*}
, where $\ell > \ell_0$, $\frac{\epsilon^2}{G_{L, \alpha, \epsilon, \ell_0}} = -C_2 \log (1 - \epsilon) - \log (-\frac{2}{\pi} \Gamma(-\alpha C_2) \Gamma(1 + C_2) \sin (\frac{\pi \alpha C_2}{2}) ) - \ell_0 C_2 \log (\frac{2}{\pi} \Gamma(\frac{\alpha}{\ell_0}) \Gamma(1 - \frac{1}{\ell_0}) \sin (\frac{\pi}{2} \frac{\alpha}{\ell_0}))$, and $C_2 = \frac{12}{\pi^2}\frac{\epsilon}{(2 + \alpha^2)}$.
\label{lem:tail-bounds}
\end{lemma}
\vspace{-1em}
With Lemma~\ref{lem:tail-bounds}, Taking $c \geq \max (G_{R, \alpha, \epsilon}, G_{L, \alpha, \epsilon, \ell_0})$ and $\delta = \exp (- \frac{\ell \epsilon^2}{c})$, we are able to prove $\mathbb{P}[\hat{s}_{\chi^2}<(1-\epsilon)s_{\chi^2} \vee \hat{s}_{\chi^2}>(1+\epsilon)s_{\chi^2}]\leq \delta$, and the above bound holds when $\ell = \frac{c}{\epsilon^2} \log (1/\delta)$. \qedhere

\end{proof}

\vspace{-1em}
\subsection{Communication \& Computation Analysis}
\label{subsec:cost-analysis}
\vspace{-0.6em}
In this section, we present the communication and computation cost of \A~\ref{alg:main}.

\begin{theorem}[Communication Cost]
Let $\Pi$ be an instantiation of \A~\ref{alg:main} with secure aggregation protocol in \A~\ref{alg:secureagg} of Appendix~\ref{sec:secureagg}, then (1) the client-side communication cost is $\mathcal{O}(\log n + m_x + m_y + \ell)$; (2) the server-side communication cost is $\mathcal{O}(n\log n + nm_x + nm_y + n\ell)$.
\label{thm:comm}
\end{theorem}


\begin{theorem}[Computation Cost]
Let $\Pi$ be an instantiation of \A~\ref{alg:main} with secure aggregation protocol in \A~\ref{alg:secureagg} of Appendix~\ref{sec:secureagg}, then (1) the client-side computation cost is 
$\mathcal{O}(m_x\log n + m_y \log n + \ell\log n + m\ell)$; 
%
(2) the server-side computation cost is 
$\mathcal{O}(m_x + m_y + \ell)$.
\label{thm:comp}
\end{theorem}


Note that compared with the original computation cost presented in~\cite{bell2020secure},
the client-side overhead has an extra $\mathcal{O}(m\ell)$ term. This term is incurred by the encoding overhead. We also give an empirical evaluation on the client-side computation overhead in \S~\ref{subsec:eval-results}. 
Please refer to Appendix~\ref{sec:proof-cost} for the detailed proof of Theorem~\ref{thm:comm} and Theorem~\ref{thm:comp}.

\vspace{-0.8em}
\section{Evaluation}
\label{sec:eval}
\vspace{-0.6em}

\paragraph{Experiment Setup.}
To benchmark \fedchi\ in terms of its accuracy drop compared with standard
$\chi^2$-test, we simulate \fedchi\ on four synthetic datasets and 16 real-world
datasets. We record the multiplicative errors of \fedchi\ compared with the
corresponding centralized $\chi^2$-test. The four synthetic datasets are
independent, linearly correlated, quadratically correlated and logistically
correlated, respectively.
As the real-world datasets, we report the details in
Appendix~\ref{sec:dataset-detail}.

We further assess the utility of \fedchi\ using three real-world downstream use
cases: 1) feature selection, 2) cryptanalysis, and 3) online false discovery
rate (FDR) control. For feature selection, we report the model accuracy trained
on the selected features. For cryptanalysis, we report the success rate of
cracking ciphertext. For Online FDR control, we report the average false
discovery rate. For all three experiments, we compare the performance of
\fedchi\ with the centralized $\chi^2$-test.

Unless otherwise specified, experiments are launched on a Ubuntu18.04 LTS server
with 32 AMD Opteron(TM) Processor 6212 with 512GB RAM.

\vspace{-0.8em}
\subsection{Evaluation Results}
\label{subsec:eval-results}
\vspace{-0.6em}

\paragraph{Accuracy.}
%
\newcommand{\figwidth}{0.24\textwidth}
\tikzset{font={\fontsize{15pt}{12}\selectfont}}
\captionsetup{font={footnotesize},skip=2pt}
\captionsetup[sub]{font={footnotesize},skip=2pt}
\begin{figure}[htb]
     \centering
     \begin{subfigure}[b]{\figwidth}
         \centering
         \resizebox{\textwidth}{!}{\begin{tikzpicture}
\begin{axis}[
  grid=major,
  ymin=0, ymax=1, xmin=10, xmax=200,
  ytick align=outside, ytick pos=left,
  xtick align=outside, xtick pos=left,
  x label style={at={(axis description cs:0.5,-0.1)},anchor=north},
  y label style={at={(axis description cs:0,.5)},anchor=south},
  ylabel={Multiplicative Error $\epsilon$},
  legend pos=north east,
  legend style={draw=none}]

\addplot+[
  blue, mark options={scale=0.75},
  smooth, 
  error bars/.cd, 
    y fixed,
    y dir=both, 
    y explicit
] table [x=x, y=y, col sep=comma] {data/basic/10/s1.txt};
\addlegendentry{$n=10$}

\addplot [name path=upper,draw=none,forget plot] table[x=x, y expr=\thisrow{y}+\thisrow{err},col sep=comma] {data/basic/10/s1.txt};
\addplot [name path=lower,draw=none,forget plot] table[x=x, y expr=\thisrow{y}-\thisrow{err},col sep=comma] {data/basic/10/s1.txt};
\addplot [fill=blue!10,fill opacity=0.8,forget plot] fill between[of=upper and lower];

\addplot+[
  green, mark options={scale=0.75},
  smooth, 
  error bars/.cd, 
    y fixed,
    y dir=both, 
    y explicit
] table [x=x, y=y, col sep=comma] {data/basic/100/s1.txt};
\addlegendentry{$n=100$}

\addplot [name path=upper,draw=none,forget plot] table[x=x, y expr=\thisrow{y}+\thisrow{err},col sep=comma] {data/basic/100/s1.txt};
\addplot [name path=lower,draw=none,forget plot] table[x=x, y expr=\thisrow{y}-\thisrow{err},col sep=comma] {data/basic/100/s1.txt};
\addplot [fill=green!10,fill opacity=0.8,forget plot] fill between[of=upper and lower];

\addplot+[
  brown, mark options={scale=0.75},
  smooth, 
  error bars/.cd, 
    y fixed,
    y dir=both, 
    y explicit
] table [x=x, y=y, col sep=comma] {data/basic/1000/s1.txt};
\addlegendentry{$n=1000$}

\addplot [name path=upper,draw=none,forget plot] table[x=x, y expr=\thisrow{y}+\thisrow{err},col sep=comma] {data/basic/1000/s1.txt};
\addplot [name path=lower,draw=none,forget plot] table[x=x, y expr=\thisrow{y}-\thisrow{err},col sep=comma] {data/basic/1000/s1.txt};
\addplot [fill=brown!10,fill opacity=0.8,forget plot] fill between[of=upper and lower];

\end{axis}
\end{tikzpicture}}
         \caption{Synthetic Data 1.}
         \label{fig:s1}
     \end{subfigure}
     \hfill     
     \begin{subfigure}[b]{\figwidth}
         \centering
         \resizebox{\textwidth}{!}{\begin{tikzpicture}
\begin{axis}[
  grid=major,
  ymin=0, ymax=1, xmin=10, xmax=200,
  ytick align=outside, ytick pos=left,
  xtick align=outside, xtick pos=left,
  x label style={at={(axis description cs:0.5,-0.1)},anchor=north},
  y label style={at={(axis description cs:0,.5)},anchor=south},
  ylabel={Multiplicative Error $\epsilon$},
  legend pos=north east,
  legend style={draw=none}]

\addplot+[
  blue, mark options={scale=0.75},
  smooth, 
  error bars/.cd, 
    y fixed,
    y dir=both, 
    y explicit
] table [x=x, y=y, col sep=comma] {data/basic/10/s2.txt};
\addlegendentry{$n=10$}

\addplot [name path=upper,draw=none,forget plot] table[x=x, y expr=\thisrow{y}+\thisrow{err},col sep=comma] {data/basic/10/s2.txt};
\addplot [name path=lower,draw=none,forget plot] table[x=x, y expr=\thisrow{y}-\thisrow{err},col sep=comma] {data/basic/10/s2.txt};
\addplot [fill=blue!10,fill opacity=0.8,forget plot] fill between[of=upper and lower];

\addplot+[
  green, mark options={scale=0.75},
  smooth, 
  error bars/.cd, 
    y fixed,
    y dir=both, 
    y explicit
] table [x=x, y=y, col sep=comma] {data/basic/100/s2.txt};
\addlegendentry{$n=100$}

\addplot [name path=upper,draw=none,forget plot] table[x=x, y expr=\thisrow{y}+\thisrow{err},col sep=comma] {data/basic/100/s2.txt};
\addplot [name path=lower,draw=none,forget plot] table[x=x, y expr=\thisrow{y}-\thisrow{err},col sep=comma] {data/basic/100/s2.txt};
\addplot [fill=green!10,fill opacity=0.8,forget plot] fill between[of=upper and lower];

\addplot+[
  brown, mark options={scale=0.75},
  smooth, 
  error bars/.cd, 
    y fixed,
    y dir=both, 
    y explicit
] table [x=x, y=y, col sep=comma] {data/basic/1000/s2.txt};
\addlegendentry{$n=1000$}

\addplot [name path=upper,draw=none,forget plot] table[x=x, y expr=\thisrow{y}+\thisrow{err},col sep=comma] {data/basic/1000/s2.txt};
\addplot [name path=lower,draw=none,forget plot] table[x=x, y expr=\thisrow{y}-\thisrow{err},col sep=comma] {data/basic/1000/s2.txt};
\addplot [fill=brown!10,fill opacity=0.8,forget plot] fill between[of=upper and lower];

\end{axis}
\end{tikzpicture}}
         \caption{Synthetic Data 2.}
         \label{fig:s2}
     \end{subfigure}
     \hfill     
     \begin{subfigure}[b]{\figwidth}
         \centering
         \resizebox{\textwidth}{!}{\begin{tikzpicture}
\begin{axis}[
  grid=major,
  ymin=0, ymax=1, xmin=10, xmax=200,
  ytick align=outside, ytick pos=left,
  xtick align=outside, xtick pos=left,
  x label style={at={(axis description cs:0.5,-0.1)},anchor=north},
  y label style={at={(axis description cs:0,.5)},anchor=south},
  ylabel={Multiplicative Error $\epsilon$},
  legend pos=north east,
  legend style={draw=none}]

\addplot+[
  blue, mark options={scale=0.75},
  smooth, 
  error bars/.cd, 
    y fixed,
    y dir=both, 
    y explicit
] table [x=x, y=y, col sep=comma] {data/basic/10/s3.txt};
\addlegendentry{$n=10$}

\addplot [name path=upper,draw=none,forget plot] table[x=x, y expr=\thisrow{y}+\thisrow{err},col sep=comma] {data/basic/10/s3.txt};
\addplot [name path=lower,draw=none,forget plot] table[x=x, y expr=\thisrow{y}-\thisrow{err},col sep=comma] {data/basic/10/s3.txt};
\addplot [fill=blue!10,fill opacity=0.8,forget plot] fill between[of=upper and lower];

\addplot+[
  green, mark options={scale=0.75},
  smooth, 
  error bars/.cd, 
    y fixed,
    y dir=both, 
    y explicit
] table [x=x, y=y, col sep=comma] {data/basic/100/s3.txt};
\addlegendentry{$n=100$}

\addplot [name path=upper,draw=none,forget plot] table[x=x, y expr=\thisrow{y}+\thisrow{err},col sep=comma] {data/basic/100/s3.txt};
\addplot [name path=lower,draw=none,forget plot] table[x=x, y expr=\thisrow{y}-\thisrow{err},col sep=comma] {data/basic/100/s3.txt};
\addplot [fill=green!10,fill opacity=0.8,forget plot] fill between[of=upper and lower];

\addplot+[
  brown, mark options={scale=0.75},
  smooth, 
  error bars/.cd, 
    y fixed,
    y dir=both, 
    y explicit
] table [x=x, y=y, col sep=comma] {data/basic/1000/s3.txt};
\addlegendentry{$n=1000$}

\addplot [name path=upper,draw=none,forget plot] table[x=x, y expr=\thisrow{y}+\thisrow{err},col sep=comma] {data/basic/1000/s3.txt};
\addplot [name path=lower,draw=none,forget plot] table[x=x, y expr=\thisrow{y}-\thisrow{err},col sep=comma] {data/basic/1000/s3.txt};
\addplot [fill=brown!10,fill opacity=0.8,forget plot] fill between[of=upper and lower];

\end{axis}
\end{tikzpicture}}
         \caption{Synthetic Data 3.}
         \label{fig:s3}
     \end{subfigure}
     \hfill     
     \begin{subfigure}[b]{\figwidth}
         \centering
         \resizebox{\textwidth}{!}{\begin{tikzpicture}
\begin{axis}[
  grid=major,
  ymin=0, ymax=1, xmin=10, xmax=200,
  ytick align=outside, ytick pos=left,
  xtick align=outside, xtick pos=left,
  x label style={at={(axis description cs:0.5,-0.1)},anchor=north},
  y label style={at={(axis description cs:0,.5)},anchor=south},
  ylabel={Multiplicative Error $\epsilon$},
  legend pos=north east,
  legend style={draw=none}]

\addplot+[
  blue, mark options={scale=0.75},
  smooth, 
  error bars/.cd, 
    y fixed,
    y dir=both, 
    y explicit
] table [x=x, y=y, col sep=comma] {data/basic/10/s4.txt};
\addlegendentry{$n=10$}

\addplot [name path=upper,draw=none,forget plot] table[x=x, y expr=\thisrow{y}+\thisrow{err},col sep=comma] {data/basic/10/s4.txt};
\addplot [name path=lower,draw=none,forget plot] table[x=x, y expr=\thisrow{y}-\thisrow{err},col sep=comma] {data/basic/10/s4.txt};
\addplot [fill=blue!10,fill opacity=0.8,forget plot] fill between[of=upper and lower];

\addplot+[
  green, mark options={scale=0.75},
  smooth, 
  error bars/.cd, 
    y fixed,
    y dir=both, 
    y explicit
] table [x=x, y=y, col sep=comma] {data/basic/100/s4.txt};
\addlegendentry{$n=100$}

\addplot [name path=upper,draw=none,forget plot] table[x=x, y expr=\thisrow{y}+\thisrow{err},col sep=comma] {data/basic/100/s4.txt};
\addplot [name path=lower,draw=none,forget plot] table[x=x, y expr=\thisrow{y}-\thisrow{err},col sep=comma] {data/basic/100/s4.txt};
\addplot [fill=green!10,fill opacity=0.8,forget plot] fill between[of=upper and lower];

\addplot+[
  brown, mark options={scale=0.75},
  smooth, 
  error bars/.cd, 
    y fixed,
    y dir=both, 
    y explicit
] table [x=x, y=y, col sep=comma] {data/basic/1000/s4.txt};
\addlegendentry{$n=1000$}

\addplot [name path=upper,draw=none,forget plot] table[x=x, y expr=\thisrow{y}+\thisrow{err},col sep=comma] {data/basic/1000/s4.txt};
\addplot [name path=lower,draw=none,forget plot] table[x=x, y expr=\thisrow{y}-\thisrow{err},col sep=comma] {data/basic/1000/s4.txt};
\addplot [fill=brown!10,fill opacity=0.8,forget plot] fill between[of=upper and lower];

\end{axis}
\end{tikzpicture}}
         \caption{Synthetic Data 4.}
         \label{fig:s4}
     \end{subfigure}
     
     \begin{subfigure}[b]{\figwidth}
         \centering
         \resizebox{\textwidth}{!}{\begin{tikzpicture}
\begin{axis}[
  grid=major,
  ymin=0, ymax=1, xmin=10, xmax=200,
  ytick align=outside, ytick pos=left,
  xtick align=outside, xtick pos=left,
  x label style={at={(axis description cs:0.5,-0.1)},anchor=north},
  y label style={at={(axis description cs:0,.5)},anchor=south},
  ylabel={Multiplicative Error $\epsilon$},
  legend pos=north east,
  legend style={draw=none}]

\addplot+[
  blue, mark options={scale=0.75},
  smooth, 
  error bars/.cd, 
    y fixed,
    y dir=both, 
    y explicit
] table [x=x, y=y, col sep=comma] {data/basic/10/adult.txt};
\addlegendentry{$n=10$}

\addplot [name path=upper,draw=none,forget plot] table[x=x, y expr=\thisrow{y}+\thisrow{err},col sep=comma] {data/basic/10/adult.txt};
\addplot [name path=lower,draw=none,forget plot] table[x=x, y expr=\thisrow{y}-\thisrow{err},col sep=comma] {data/basic/10/adult.txt};
\addplot [fill=blue!10,fill opacity=0.8,forget plot] fill between[of=upper and lower];

\addplot+[
  green, mark options={scale=0.75},
  smooth, 
  error bars/.cd, 
    y fixed,
    y dir=both, 
    y explicit
] table [x=x, y=y, col sep=comma] {data/basic/100/adult.txt};
\addlegendentry{$n=100$}

\addplot [name path=upper,draw=none,forget plot] table[x=x, y expr=\thisrow{y}+\thisrow{err},col sep=comma] {data/basic/100/adult.txt};
\addplot [name path=lower,draw=none,forget plot] table[x=x, y expr=\thisrow{y}-\thisrow{err},col sep=comma] {data/basic/100/adult.txt};
\addplot [fill=green!10,fill opacity=0.8,forget plot] fill between[of=upper and lower];

\addplot+[
  brown, mark options={scale=0.75},
  smooth, 
  error bars/.cd, 
    y fixed,
    y dir=both, 
    y explicit
] table [x=x, y=y, col sep=comma] {data/basic/1000/adult.txt};
\addlegendentry{$n=1000$}

\addplot [name path=upper,draw=none,forget plot] table[x=x, y expr=\thisrow{y}+\thisrow{err},col sep=comma] {data/basic/1000/adult.txt};
\addplot [name path=lower,draw=none,forget plot] table[x=x, y expr=\thisrow{y}-\thisrow{err},col sep=comma] {data/basic/1000/adult.txt};
\addplot [fill=brown!10,fill opacity=0.8,forget plot] fill between[of=upper and lower];

\end{axis}
\end{tikzpicture}}
         \caption{Data 1.}
         \label{fig:d1}
     \end{subfigure}
     \hfill     
     \begin{subfigure}[b]{\figwidth}
         \centering
         \resizebox{\textwidth}{!}{\begin{tikzpicture}
\begin{axis}[
  grid=major,
  ymin=0, ymax=1, xmin=10, xmax=100,
  ytick align=outside, ytick pos=left,
  xtick align=outside, xtick pos=left,
  x label style={at={(axis description cs:0.5,-0.1)},anchor=north},
  y label style={at={(axis description cs:0,.5)},anchor=south},
  ylabel={Multiplicative Error $\epsilon$},
  legend pos=north east,
  legend style={draw=none}]
  
\addplot+[
  blue, mark options={scale=0.75},
  smooth, 
  error bars/.cd, 
    y fixed,
    y dir=both, 
    y explicit
] table [x=x, y=y, col sep=comma] {data/basic/10/credit_1.txt};
\addlegendentry{$n=10$}

\addplot [name path=upper,draw=none,forget plot] table[x=x, y expr=\thisrow{y}+\thisrow{err},col sep=comma] {data/basic/10/credit_1.txt};
\addplot [name path=lower,draw=none,forget plot] table[x=x, y expr=\thisrow{y}-\thisrow{err},col sep=comma] {data/basic/10/credit_1.txt};
\addplot [fill=blue!10,fill opacity=0.8,forget plot] fill between[of=upper and lower];

\addplot+[
  green, mark options={scale=0.75},
  smooth, 
  error bars/.cd, 
    y fixed,
    y dir=both, 
    y explicit
] table [x=x, y=y, col sep=comma] {data/basic/100/credit_1.txt};
\addlegendentry{$n=100$}

\addplot [name path=upper,draw=none,forget plot] table[x=x, y expr=\thisrow{y}+\thisrow{err},col sep=comma] {data/basic/100/credit_1.txt};
\addplot [name path=lower,draw=none,forget plot] table[x=x, y expr=\thisrow{y}-\thisrow{err},col sep=comma] {data/basic/100/credit_1.txt};
\addplot [fill=green!10,fill opacity=0.8,forget plot] fill between[of=upper and lower];

\addplot+[
  brown, mark options={scale=0.75},
  smooth, 
  error bars/.cd, 
    y fixed,
    y dir=both, 
    y explicit
] table [x=x, y=y, col sep=comma] {data/basic/1000/credit_1.txt};
\addlegendentry{$n=1000$}

\addplot [name path=upper,draw=none,forget plot] table[x=x, y expr=\thisrow{y}+\thisrow{err},col sep=comma] {data/basic/1000/credit_1.txt};
\addplot [name path=lower,draw=none,forget plot] table[x=x, y expr=\thisrow{y}-\thisrow{err},col sep=comma] {data/basic/1000/credit_1.txt};
\addplot [fill=brown!10,fill opacity=0.8,forget plot] fill between[of=upper and lower];

\end{axis}
\end{tikzpicture}}
         \caption{Data 2.}
         \label{fig:d2}
     \end{subfigure}
     \hfill     
     \begin{subfigure}[b]{\figwidth}
         \centering
         \resizebox{\textwidth}{!}{\begin{tikzpicture}
\begin{axis}[
  grid=major,
  ymin=0, ymax=1, xmin=10, xmax=200,
  ytick align=outside, ytick pos=left,
  xtick align=outside, xtick pos=left,
  x label style={at={(axis description cs:0.5,-0.1)},anchor=north},
  y label style={at={(axis description cs:0,.5)},anchor=south},
  ylabel={Multiplicative Error $\epsilon$},
  legend pos=north east,
  legend style={draw=none}]
  
\addplot+[
  blue, mark options={scale=0.75},
  smooth, 
  error bars/.cd, 
    y fixed,
    y dir=both, 
    y explicit
] table [x=x, y=y, col sep=comma] {data/basic/10/credit_2.txt};
\addlegendentry{$n=10$}

\addplot [name path=upper,draw=none,forget plot] table[x=x, y expr=\thisrow{y}+\thisrow{err},col sep=comma] {data/basic/10/credit_2.txt};
\addplot [name path=lower,draw=none,forget plot] table[x=x, y expr=\thisrow{y}-\thisrow{err},col sep=comma] {data/basic/10/credit_2.txt};
\addplot [fill=blue!10,fill opacity=0.8,forget plot] fill between[of=upper and lower];

\addplot+[
  green, mark options={scale=0.75},
  smooth, 
  error bars/.cd, 
    y fixed,
    y dir=both, 
    y explicit
] table [x=x, y=y, col sep=comma] {data/basic/100/credit_2.txt};
\addlegendentry{$n=100$}

\addplot [name path=upper,draw=none,forget plot] table[x=x, y expr=\thisrow{y}+\thisrow{err},col sep=comma] {data/basic/100/credit_2.txt};
\addplot [name path=lower,draw=none,forget plot] table[x=x, y expr=\thisrow{y}-\thisrow{err},col sep=comma] {data/basic/100/credit_2.txt};
\addplot [fill=green!10,fill opacity=0.8,forget plot] fill between[of=upper and lower];

\addplot+[
  brown, mark options={scale=0.75},
  smooth, 
  error bars/.cd, 
    y fixed,
    y dir=both, 
    y explicit
] table [x=x, y=y, col sep=comma] {data/basic/1000/credit_2.txt};
\addlegendentry{$n=1000$}

\addplot [name path=upper,draw=none,forget plot] table[x=x, y expr=\thisrow{y}+\thisrow{err},col sep=comma] {data/basic/1000/credit_2.txt};
\addplot [name path=lower,draw=none,forget plot] table[x=x, y expr=\thisrow{y}-\thisrow{err},col sep=comma] {data/basic/1000/credit_2.txt};
\addplot [fill=brown!10,fill opacity=0.8,forget plot] fill between[of=upper and lower];

\end{axis}
\end{tikzpicture}}
         \caption{Data 3.}
         \label{fig:d3}
     \end{subfigure}
     \hfill     
     \begin{subfigure}[b]{\figwidth}
         \centering
         \resizebox{\textwidth}{!}{\begin{tikzpicture}
\begin{axis}[
  grid=major,
  ymin=0, ymax=1, xmin=10, xmax=200,
  ytick align=outside, ytick pos=left,
  xtick align=outside, xtick pos=left,
  x label style={at={(axis description cs:0.5,-0.1)},anchor=north},
  y label style={at={(axis description cs:0,.5)},anchor=south},
  ylabel={Multiplicative Error $\epsilon$},
  legend pos=north east,
  legend style={draw=none}]
  
\addplot+[
  blue, mark options={scale=0.75},
  smooth, 
  error bars/.cd, 
    y fixed,
    y dir=both, 
    y explicit
] table [x=x, y=y, col sep=comma] {data/basic/10/credit_3.txt};
\addlegendentry{$n=10$}

\addplot [name path=upper,draw=none,forget plot] table[x=x, y expr=\thisrow{y}+\thisrow{err},col sep=comma] {data/basic/10/credit_3.txt};
\addplot [name path=lower,draw=none,forget plot] table[x=x, y expr=\thisrow{y}-\thisrow{err},col sep=comma] {data/basic/10/credit_3.txt};
\addplot [fill=blue!10,fill opacity=0.8,forget plot] fill between[of=upper and lower];

\addplot+[
  green, mark options={scale=0.75},
  smooth, 
  error bars/.cd, 
    y fixed,
    y dir=both, 
    y explicit
] table [x=x, y=y, col sep=comma] {data/basic/100/credit_3.txt};
\addlegendentry{$n=100$}

\addplot [name path=upper,draw=none,forget plot] table[x=x, y expr=\thisrow{y}+\thisrow{err},col sep=comma] {data/basic/100/credit_3.txt};
\addplot [name path=lower,draw=none,forget plot] table[x=x, y expr=\thisrow{y}-\thisrow{err},col sep=comma] {data/basic/100/credit_3.txt};
\addplot [fill=green!10,fill opacity=0.8,forget plot] fill between[of=upper and lower];

\addplot+[
  brown, mark options={scale=0.75},
  smooth, 
  error bars/.cd, 
    y fixed,
    y dir=both, 
    y explicit
] table [x=x, y=y, col sep=comma] {data/basic/1000/credit_3.txt};
\addlegendentry{$n=1000$}

\addplot [name path=upper,draw=none,forget plot] table[x=x, y expr=\thisrow{y}+\thisrow{err},col sep=comma] {data/basic/1000/credit_3.txt};
\addplot [name path=lower,draw=none,forget plot] table[x=x, y expr=\thisrow{y}-\thisrow{err},col sep=comma] {data/basic/1000/credit_3.txt};
\addplot [fill=brown!10,fill opacity=0.8,forget plot] fill between[of=upper and lower];

\end{axis}
\end{tikzpicture}}
         \caption{Data 4.}
         \label{fig:d4}
     \end{subfigure}
     
     \begin{subfigure}[b]{\figwidth}
         \centering
         \resizebox{\textwidth}{!}{\begin{tikzpicture}
\begin{axis}[
  grid=major,
  ymin=0, ymax=1, xmin=10, xmax=200,
  ytick align=outside, ytick pos=left,
  xtick align=outside, xtick pos=left,
  x label style={at={(axis description cs:0.5,-0.1)},anchor=north},
  y label style={at={(axis description cs:0,.5)},anchor=south},
  ylabel={Multiplicative Error $\epsilon$},
  legend pos=north east,
  legend style={draw=none}]
  
\addplot+[
  blue, mark options={scale=0.75},
  smooth, 
  error bars/.cd, 
    y fixed,
    y dir=both, 
    y explicit
] table [x=x, y=y, col sep=comma] {data/basic/10/credit_4.txt};
\addlegendentry{$n=10$}

\addplot [name path=upper,draw=none,forget plot] table[x=x, y expr=\thisrow{y}+\thisrow{err},col sep=comma] {data/basic/10/credit_4.txt};
\addplot [name path=lower,draw=none,forget plot] table[x=x, y expr=\thisrow{y}-\thisrow{err},col sep=comma] {data/basic/10/credit_4.txt};
\addplot [fill=blue!10,fill opacity=0.8,forget plot] fill between[of=upper and lower];

\addplot+[
  green, mark options={scale=0.75},
  smooth, 
  error bars/.cd, 
    y fixed,
    y dir=both, 
    y explicit
] table [x=x, y=y, col sep=comma] {data/basic/100/credit_4.txt};
\addlegendentry{$n=100$}

\addplot [name path=upper,draw=none,forget plot] table[x=x, y expr=\thisrow{y}+\thisrow{err},col sep=comma] {data/basic/100/credit_4.txt};
\addplot [name path=lower,draw=none,forget plot] table[x=x, y expr=\thisrow{y}-\thisrow{err},col sep=comma] {data/basic/100/credit_4.txt};
\addplot [fill=green!10,fill opacity=0.8,forget plot] fill between[of=upper and lower];

\addplot+[
  brown, mark options={scale=0.75},
  smooth, 
  error bars/.cd, 
    y fixed,
    y dir=both, 
    y explicit
] table [x=x, y=y, col sep=comma] {data/basic/1000/credit_4.txt};
\addlegendentry{$n=1000$}

\addplot [name path=upper,draw=none,forget plot] table[x=x, y expr=\thisrow{y}+\thisrow{err},col sep=comma] {data/basic/1000/credit_4.txt};
\addplot [name path=lower,draw=none,forget plot] table[x=x, y expr=\thisrow{y}-\thisrow{err},col sep=comma] {data/basic/1000/credit_4.txt};
\addplot [fill=brown!10,fill opacity=0.8,forget plot] fill between[of=upper and lower];

\end{axis}
\end{tikzpicture}}
         \caption{Data 5.}
         \label{fig:d5}
     \end{subfigure}
     \hfill     
     \begin{subfigure}[b]{\figwidth}
         \centering
         \resizebox{\textwidth}{!}{\begin{tikzpicture}
\begin{axis}[
  grid=major,
  ymin=0, ymax=1, xmin=10, xmax=200,
  ytick align=outside, ytick pos=left,
  xtick align=outside, xtick pos=left,
  x label style={at={(axis description cs:0.5,-0.1)},anchor=north},
  y label style={at={(axis description cs:0,.5)},anchor=south},
  ylabel={Multiplicative Error $\epsilon$},
  legend pos=north east,
  legend style={draw=none}]
  
\addplot+[
  blue, mark options={scale=0.75},
  smooth, 
  error bars/.cd, 
    y fixed,
    y dir=both, 
    y explicit
] table [x=x, y=y, col sep=comma] {data/basic/10/gtsrb_1.txt};
\addlegendentry{$n=10$}

\addplot [name path=upper,draw=none,forget plot] table[x=x, y expr=\thisrow{y}+\thisrow{err},col sep=comma] {data/basic/10/gtsrb_1.txt};
\addplot [name path=lower,draw=none,forget plot] table[x=x, y expr=\thisrow{y}-\thisrow{err},col sep=comma] {data/basic/10/gtsrb_1.txt};
\addplot [fill=blue!10,fill opacity=0.8,forget plot] fill between[of=upper and lower];

\addplot+[
  green, mark options={scale=0.75},
  smooth, 
  error bars/.cd, 
    y fixed,
    y dir=both, 
    y explicit
] table [x=x, y=y, col sep=comma] {data/basic/100/gtsrb_1.txt};
\addlegendentry{$n=100$}

\addplot [name path=upper,draw=none,forget plot] table[x=x, y expr=\thisrow{y}+\thisrow{err},col sep=comma] {data/basic/100/gtsrb_1.txt};
\addplot [name path=lower,draw=none,forget plot] table[x=x, y expr=\thisrow{y}-\thisrow{err},col sep=comma] {data/basic/100/gtsrb_1.txt};
\addplot [fill=green!10,fill opacity=0.8,forget plot] fill between[of=upper and lower];

\addplot+[
  brown, mark options={scale=0.75},
  smooth, 
  error bars/.cd, 
    y fixed,
    y dir=both, 
    y explicit
] table [x=x, y=y, col sep=comma] {data/basic/1000/gtsrb_1.txt};
\addlegendentry{$n=1000$}

\addplot [name path=upper,draw=none,forget plot] table[x=x, y expr=\thisrow{y}+\thisrow{err},col sep=comma] {data/basic/1000/gtsrb_1.txt};
\addplot [name path=lower,draw=none,forget plot] table[x=x, y expr=\thisrow{y}-\thisrow{err},col sep=comma] {data/basic/1000/gtsrb_1.txt};
\addplot [fill=brown!10,fill opacity=0.8,forget plot] fill between[of=upper and lower];

\end{axis}
\end{tikzpicture}}
         \caption{Data 6.}
         \label{fig:d6}
     \end{subfigure}
     \hfill     
     \begin{subfigure}[b]{\figwidth}
         \centering
         \resizebox{\textwidth}{!}{\begin{tikzpicture}
\begin{axis}[
  grid=major,
  ymin=0, ymax=1, xmin=10, xmax=200,
  ytick align=outside, ytick pos=left,
  xtick align=outside, xtick pos=left,
  x label style={at={(axis description cs:0.5,-0.1)},anchor=north},
  y label style={at={(axis description cs:0,.5)},anchor=south},
  ylabel={Multiplicative Error $\epsilon$},
  legend pos=north east,
  legend style={draw=none}]
  
\addplot+[
  blue, mark options={scale=0.75},
  smooth, 
  error bars/.cd, 
    y fixed,
    y dir=both, 
    y explicit
] table [x=x, y=y, col sep=comma] {data/basic/10/gtsrb_2.txt};
\addlegendentry{$n=10$}

\addplot [name path=upper,draw=none,forget plot] table[x=x, y expr=\thisrow{y}+\thisrow{err},col sep=comma] {data/basic/10/gtsrb_2.txt};
\addplot [name path=lower,draw=none,forget plot] table[x=x, y expr=\thisrow{y}-\thisrow{err},col sep=comma] {data/basic/10/gtsrb_2.txt};
\addplot [fill=blue!10,fill opacity=0.8,forget plot] fill between[of=upper and lower];

\addplot+[
  green, mark options={scale=0.75},
  smooth, 
  error bars/.cd, 
    y fixed,
    y dir=both, 
    y explicit
] table [x=x, y=y, col sep=comma] {data/basic/100/gtsrb_2.txt};
\addlegendentry{$n=100$}

\addplot [name path=upper,draw=none,forget plot] table[x=x, y expr=\thisrow{y}+\thisrow{err},col sep=comma] {data/basic/100/gtsrb_2.txt};
\addplot [name path=lower,draw=none,forget plot] table[x=x, y expr=\thisrow{y}-\thisrow{err},col sep=comma] {data/basic/100/gtsrb_2.txt};
\addplot [fill=green!10,fill opacity=0.8,forget plot] fill between[of=upper and lower];

\addplot+[
  brown, mark options={scale=0.75},
  smooth, 
  error bars/.cd, 
    y fixed,
    y dir=both, 
    y explicit
] table [x=x, y=y, col sep=comma] {data/basic/1000/gtsrb_2.txt};
\addlegendentry{$n=1000$}

\addplot [name path=upper,draw=none,forget plot] table[x=x, y expr=\thisrow{y}+\thisrow{err},col sep=comma] {data/basic/1000/gtsrb_2.txt};
\addplot [name path=lower,draw=none,forget plot] table[x=x, y expr=\thisrow{y}-\thisrow{err},col sep=comma] {data/basic/1000/gtsrb_2.txt};
\addplot [fill=brown!10,fill opacity=0.8,forget plot] fill between[of=upper and lower];

\end{axis}
\end{tikzpicture}}
         \caption{Data 7.}
         \label{fig:d7}
     \end{subfigure}
     \hfill     
     \begin{subfigure}[b]{\figwidth}
         \centering
         \resizebox{\textwidth}{!}{\begin{tikzpicture}
\begin{axis}[
  grid=major,
  ymin=0, ymax=1, xmin=10, xmax=200,
  ytick align=outside, ytick pos=left,
  xtick align=outside, xtick pos=left,
  x label style={at={(axis description cs:0.5,-0.1)},anchor=north},
  y label style={at={(axis description cs:0,.5)},anchor=south},
  ylabel={Multiplicative Error $\epsilon$},
  legend pos=north east,
  legend style={draw=none}]
  
\addplot+[
  blue, mark options={scale=0.75},
  smooth, 
  error bars/.cd, 
    y fixed,
    y dir=both, 
    y explicit
] table [x=x, y=y, col sep=comma] {data/basic/10/gtsrb_3.txt};
\addlegendentry{$n=10$}

\addplot [name path=upper,draw=none,forget plot] table[x=x, y expr=\thisrow{y}+\thisrow{err},col sep=comma] {data/basic/10/gtsrb_3.txt};
\addplot [name path=lower,draw=none,forget plot] table[x=x, y expr=\thisrow{y}-\thisrow{err},col sep=comma] {data/basic/10/gtsrb_3.txt};
\addplot [fill=blue!10,fill opacity=0.8,forget plot] fill between[of=upper and lower];

\addplot+[
  green, mark options={scale=0.75},
  smooth, 
  error bars/.cd, 
    y fixed,
    y dir=both, 
    y explicit
] table [x=x, y=y, col sep=comma] {data/basic/100/gtsrb_3.txt};
\addlegendentry{$n=100$}

\addplot [name path=upper,draw=none,forget plot] table[x=x, y expr=\thisrow{y}+\thisrow{err},col sep=comma] {data/basic/100/gtsrb_3.txt};
\addplot [name path=lower,draw=none,forget plot] table[x=x, y expr=\thisrow{y}-\thisrow{err},col sep=comma] {data/basic/100/gtsrb_3.txt};
\addplot [fill=green!10,fill opacity=0.8,forget plot] fill between[of=upper and lower];

\addplot+[
  brown, mark options={scale=0.75},
  smooth, 
  error bars/.cd, 
    y fixed,
    y dir=both, 
    y explicit
] table [x=x, y=y, col sep=comma] {data/basic/1000/gtsrb_3.txt};
\addlegendentry{$n=1000$}

\addplot [name path=upper,draw=none,forget plot] table[x=x, y expr=\thisrow{y}+\thisrow{err},col sep=comma] {data/basic/1000/gtsrb_3.txt};
\addplot [name path=lower,draw=none,forget plot] table[x=x, y expr=\thisrow{y}-\thisrow{err},col sep=comma] {data/basic/1000/gtsrb_3.txt};
\addplot [fill=brown!10,fill opacity=0.8,forget plot] fill between[of=upper and lower];

\end{axis}
\end{tikzpicture}}
         \caption{Data 8.}
         \label{fig:d8}
     \end{subfigure}
     
     \begin{subfigure}[b]{\figwidth}
         \centering
         \resizebox{\textwidth}{!}{\begin{tikzpicture}
\begin{axis}[
  grid=major,
  ymin=0, ymax=1, xmin=10, xmax=200,
  ytick align=outside, ytick pos=left,
  xtick align=outside, xtick pos=left,
  x label style={at={(axis description cs:0.5,-0.1)},anchor=north},
  y label style={at={(axis description cs:0,.5)},anchor=south},
  ylabel={Multiplicative Error $\epsilon$},
  legend pos=north east,
  legend style={draw=none}]
  
\addplot+[
  blue, mark options={scale=0.75},
  smooth, 
  error bars/.cd, 
    y fixed,
    y dir=both, 
    y explicit
] table [x=x, y=y, col sep=comma] {data/basic/10/gtsrb_3.txt};
\addlegendentry{$n=10$}

\addplot [name path=upper,draw=none,forget plot] table[x=x, y expr=\thisrow{y}+\thisrow{err},col sep=comma] {data/basic/10/gtsrb_4.txt};
\addplot [name path=lower,draw=none,forget plot] table[x=x, y expr=\thisrow{y}-\thisrow{err},col sep=comma] {data/basic/10/gtsrb_4.txt};
\addplot [fill=blue!10,fill opacity=0.8,forget plot] fill between[of=upper and lower];

\addplot+[
  green, mark options={scale=0.75},
  smooth, 
  error bars/.cd, 
    y fixed,
    y dir=both, 
    y explicit
] table [x=x, y=y, col sep=comma] {data/basic/100/gtsrb_3.txt};
\addlegendentry{$n=100$}

\addplot [name path=upper,draw=none,forget plot] table[x=x, y expr=\thisrow{y}+\thisrow{err},col sep=comma] {data/basic/100/gtsrb_4.txt};
\addplot [name path=lower,draw=none,forget plot] table[x=x, y expr=\thisrow{y}-\thisrow{err},col sep=comma] {data/basic/100/gtsrb_4.txt};
\addplot [fill=green!10,fill opacity=0.8,forget plot] fill between[of=upper and lower];

\addplot+[
  brown, mark options={scale=0.75},
  smooth, 
  error bars/.cd, 
    y fixed,
    y dir=both, 
    y explicit
] table [x=x, y=y, col sep=comma] {data/basic/1000/gtsrb_3.txt};
\addlegendentry{$n=1000$}

\addplot [name path=upper,draw=none,forget plot] table[x=x, y expr=\thisrow{y}+\thisrow{err},col sep=comma] {data/basic/1000/gtsrb_4.txt};
\addplot [name path=lower,draw=none,forget plot] table[x=x, y expr=\thisrow{y}-\thisrow{err},col sep=comma] {data/basic/1000/gtsrb_4.txt};
\addplot [fill=brown!10,fill opacity=0.8,forget plot] fill between[of=upper and lower];

\end{axis}
\end{tikzpicture}}
         \caption{Data 9.}
         \label{fig:d9}
     \end{subfigure}
     \hfill     
     \begin{subfigure}[b]{\figwidth}
         \centering
         \resizebox{\textwidth}{!}{\begin{tikzpicture}
\begin{axis}[
  grid=major,
  ymin=0, ymax=1, xmin=10, xmax=200,
  ytick align=outside, ytick pos=left,
  xtick align=outside, xtick pos=left,
  x label style={at={(axis description cs:0.5,-0.1)},anchor=north},
  y label style={at={(axis description cs:0,.5)},anchor=south},
  ylabel={Multiplicative Error $\epsilon$},
  legend pos=north east,
  legend style={draw=none}]
  
\addplot+[
  blue, mark options={scale=0.75},
  smooth, 
  error bars/.cd, 
    y fixed,
    y dir=both, 
    y explicit
] table [x=x, y=y, col sep=comma] {data/basic/10/gtsrb_5.txt};
\addlegendentry{$n=10$}

\addplot [name path=upper,draw=none,forget plot] table[x=x, y expr=\thisrow{y}+\thisrow{err},col sep=comma] {data/basic/10/gtsrb_5.txt};
\addplot [name path=lower,draw=none,forget plot] table[x=x, y expr=\thisrow{y}-\thisrow{err},col sep=comma] {data/basic/10/gtsrb_5.txt};
\addplot [fill=blue!10,fill opacity=0.8,forget plot] fill between[of=upper and lower];

\addplot+[
  green, mark options={scale=0.75},
  smooth, 
  error bars/.cd, 
    y fixed,
    y dir=both, 
    y explicit
] table [x=x, y=y, col sep=comma] {data/basic/100/gtsrb_5.txt};
\addlegendentry{$n=100$}

\addplot [name path=upper,draw=none,forget plot] table[x=x, y expr=\thisrow{y}+\thisrow{err},col sep=comma] {data/basic/100/gtsrb_5.txt};
\addplot [name path=lower,draw=none,forget plot] table[x=x, y expr=\thisrow{y}-\thisrow{err},col sep=comma] {data/basic/100/gtsrb_5.txt};
\addplot [fill=green!10,fill opacity=0.8,forget plot] fill between[of=upper and lower];

\addplot+[
  brown, mark options={scale=0.75},
  smooth, 
  error bars/.cd, 
    y fixed,
    y dir=both, 
    y explicit
] table [x=x, y=y, col sep=comma] {data/basic/1000/gtsrb_5.txt};
\addlegendentry{$n=1000$}

\addplot [name path=upper,draw=none,forget plot] table[x=x, y expr=\thisrow{y}+\thisrow{err},col sep=comma] {data/basic/1000/gtsrb_5.txt};
\addplot [name path=lower,draw=none,forget plot] table[x=x, y expr=\thisrow{y}-\thisrow{err},col sep=comma] {data/basic/1000/gtsrb_5.txt};
\addplot [fill=brown!10,fill opacity=0.8,forget plot] fill between[of=upper and lower];

\end{axis}
\end{tikzpicture}}
         \caption{Data 10.}
         \label{fig:d10}
     \end{subfigure}
     \hfill     
     \begin{subfigure}[b]{\figwidth}
         \centering
         \resizebox{\textwidth}{!}{\begin{tikzpicture}
\begin{axis}[
  grid=major,
  ymin=0, ymax=1, xmin=10, xmax=200,
  ytick align=outside, ytick pos=left,
  xtick align=outside, xtick pos=left,
  x label style={at={(axis description cs:0.5,-0.1)},anchor=north},
  y label style={at={(axis description cs:0,.5)},anchor=south},
  ylabel={Multiplicative Error $\epsilon$},
  legend pos=north east,
  legend style={draw=none}]
  
\addplot+[
  blue, mark options={scale=0.75},
  smooth, 
  error bars/.cd, 
    y fixed,
    y dir=both, 
    y explicit
] table [x=x, y=y, col sep=comma] {data/basic/10/gtsrb_6.txt};
\addlegendentry{$n=10$}

\addplot [name path=upper,draw=none,forget plot] table[x=x, y expr=\thisrow{y}+\thisrow{err},col sep=comma] {data/basic/10/gtsrb_6.txt};
\addplot [name path=lower,draw=none,forget plot] table[x=x, y expr=\thisrow{y}-\thisrow{err},col sep=comma] {data/basic/10/gtsrb_6.txt};
\addplot [fill=blue!10,fill opacity=0.8,forget plot] fill between[of=upper and lower];

\addplot+[
  green, mark options={scale=0.75},
  smooth, 
  error bars/.cd, 
    y fixed,
    y dir=both, 
    y explicit
] table [x=x, y=y, col sep=comma] {data/basic/100/gtsrb_6.txt};
\addlegendentry{$n=100$}

\addplot [name path=upper,draw=none,forget plot] table[x=x, y expr=\thisrow{y}+\thisrow{err},col sep=comma] {data/basic/100/gtsrb_6.txt};
\addplot [name path=lower,draw=none,forget plot] table[x=x, y expr=\thisrow{y}-\thisrow{err},col sep=comma] {data/basic/100/gtsrb_6.txt};
\addplot [fill=green!10,fill opacity=0.8,forget plot] fill between[of=upper and lower];

\addplot+[
  brown, mark options={scale=0.75},
  smooth, 
  error bars/.cd, 
    y fixed,
    y dir=both, 
    y explicit
] table [x=x, y=y, col sep=comma] {data/basic/1000/gtsrb_6.txt};
\addlegendentry{$n=1000$}

\addplot [name path=upper,draw=none,forget plot] table[x=x, y expr=\thisrow{y}+\thisrow{err},col sep=comma] {data/basic/1000/gtsrb_6.txt};
\addplot [name path=lower,draw=none,forget plot] table[x=x, y expr=\thisrow{y}-\thisrow{err},col sep=comma] {data/basic/1000/gtsrb_6.txt};
\addplot [fill=brown!10,fill opacity=0.8,forget plot] fill between[of=upper and lower];

\end{axis}
\end{tikzpicture}}
         \caption{Data 11.}
         \label{fig:d11}
     \end{subfigure}
     \hfill     
     \begin{subfigure}[b]{\figwidth}
         \centering
         \resizebox{\textwidth}{!}{\begin{tikzpicture}
\begin{axis}[
  grid=major,
  ymin=0, ymax=1, xmin=10, xmax=50,
  ytick align=outside, ytick pos=left,
  xtick align=outside, xtick pos=left,
  x label style={at={(axis description cs:0.5,-0.1)},anchor=north},
  y label style={at={(axis description cs:0,.5)},anchor=south},
  ylabel={Multiplicative Error $\epsilon$},
  legend pos=north east,
  legend style={draw=none}]
  
\addplot+[
  blue, mark options={scale=0.75},
  smooth, 
  error bars/.cd, 
    y fixed,
    y dir=both, 
    y explicit
] table [x=x, y=y, col sep=comma] {data/basic/10/mushroom_1.txt};
\addlegendentry{$n=10$}

\addplot [name path=upper,draw=none,forget plot] table[x=x, y expr=\thisrow{y}+\thisrow{err},col sep=comma] {data/basic/10/mushroom_1.txt};
\addplot [name path=lower,draw=none,forget plot] table[x=x, y expr=\thisrow{y}-\thisrow{err},col sep=comma] {data/basic/10/mushroom_1.txt};
\addplot [fill=blue!10,fill opacity=0.8,forget plot] fill between[of=upper and lower];

\addplot+[
  green, mark options={scale=0.75},
  smooth, 
  error bars/.cd, 
    y fixed,
    y dir=both, 
    y explicit
] table [x=x, y=y, col sep=comma] {data/basic/100/mushroom_1.txt};
\addlegendentry{$n=100$}

\addplot [name path=upper,draw=none,forget plot] table[x=x, y expr=\thisrow{y}+\thisrow{err},col sep=comma] {data/basic/100/mushroom_1.txt};
\addplot [name path=lower,draw=none,forget plot] table[x=x, y expr=\thisrow{y}-\thisrow{err},col sep=comma] {data/basic/100/mushroom_1.txt};
\addplot [fill=green!10,fill opacity=0.8,forget plot] fill between[of=upper and lower];

\addplot+[
  brown, mark options={scale=0.75},
  smooth, 
  error bars/.cd, 
    y fixed,
    y dir=both, 
    y explicit
] table [x=x, y=y, col sep=comma] {data/basic/1000/mushroom_1.txt};
\addlegendentry{$n=1000$}

\addplot [name path=upper,draw=none,forget plot] table[x=x, y expr=\thisrow{y}+\thisrow{err},col sep=comma] {data/basic/1000/mushroom_1.txt};
\addplot [name path=lower,draw=none,forget plot] table[x=x, y expr=\thisrow{y}-\thisrow{err},col sep=comma] {data/basic/1000/mushroom_1.txt};
\addplot [fill=brown!10,fill opacity=0.8,forget plot] fill between[of=upper and lower];

\end{axis}
\end{tikzpicture}}
         \caption{Data 12.}
         \label{fig:d12}
     \end{subfigure}
     
     \begin{subfigure}[b]{\figwidth}
         \centering
         \resizebox{\textwidth}{!}{\begin{tikzpicture}
\begin{axis}[
  grid=major,
  ymin=0, ymax=1, xmin=10, xmax=50,
  ytick align=outside, ytick pos=left,
  xtick align=outside, xtick pos=left,
  x label style={at={(axis description cs:0.5,-0.1)},anchor=north},
  y label style={at={(axis description cs:0,.5)},anchor=south},
  xlabel=Encoding size $\ell$,
  ylabel={Multiplicative Error $\epsilon$},
  legend pos=north east,
  legend style={draw=none}]
  
\addplot+[
  blue, mark options={scale=0.75},
  smooth, 
  error bars/.cd, 
    y fixed,
    y dir=both, 
    y explicit
] table [x=x, y=y, col sep=comma] {data/basic/10/mushroom_2.txt};
\addlegendentry{$n=10$}

\addplot [name path=upper,draw=none,forget plot] table[x=x, y expr=\thisrow{y}+\thisrow{err},col sep=comma] {data/basic/10/mushroom_2.txt};
\addplot [name path=lower,draw=none,forget plot] table[x=x, y expr=\thisrow{y}-\thisrow{err},col sep=comma] {data/basic/10/mushroom_2.txt};
\addplot [fill=blue!10,fill opacity=0.8,forget plot] fill between[of=upper and lower];

\addplot+[
  green, mark options={scale=0.75},
  smooth, 
  error bars/.cd, 
    y fixed,
    y dir=both, 
    y explicit
] table [x=x, y=y, col sep=comma] {data/basic/100/mushroom_2.txt};
\addlegendentry{$n=100$}

\addplot [name path=upper,draw=none,forget plot] table[x=x, y expr=\thisrow{y}+\thisrow{err},col sep=comma] {data/basic/100/mushroom_2.txt};
\addplot [name path=lower,draw=none,forget plot] table[x=x, y expr=\thisrow{y}-\thisrow{err},col sep=comma] {data/basic/100/mushroom_2.txt};
\addplot [fill=green!10,fill opacity=0.8,forget plot] fill between[of=upper and lower];

\addplot+[
  brown, mark options={scale=0.75},
  smooth, 
  error bars/.cd, 
    y fixed,
    y dir=both, 
    y explicit
] table [x=x, y=y, col sep=comma] {data/basic/1000/mushroom_2.txt};
\addlegendentry{$n=1000$}

\addplot [name path=upper,draw=none,forget plot] table[x=x, y expr=\thisrow{y}+\thisrow{err},col sep=comma] {data/basic/1000/mushroom_2.txt};
\addplot [name path=lower,draw=none,forget plot] table[x=x, y expr=\thisrow{y}-\thisrow{err},col sep=comma] {data/basic/1000/mushroom_2.txt};
\addplot [fill=brown!10,fill opacity=0.8,forget plot] fill between[of=upper and lower];

\end{axis}
\end{tikzpicture}}
         \caption{Data 13.}
         \label{fig:d13}
     \end{subfigure}
     \hfill     
     \begin{subfigure}[b]{\figwidth}
         \centering
         \resizebox{\textwidth}{!}{\begin{tikzpicture}
\begin{axis}[
  grid=major,
  ymin=0, ymax=1, xmin=10, xmax=50,
  ytick align=outside, ytick pos=left,
  xtick align=outside, xtick pos=left,
  x label style={at={(axis description cs:0.5,-0.1)},anchor=north},
  y label style={at={(axis description cs:0,.5)},anchor=south},
  xlabel=Encoding size $\ell$,
  ylabel={Multiplicative Error $\epsilon$},
  legend pos=north east,
  legend style={draw=none}]
  
\addplot+[
  blue, mark options={scale=0.75},
  smooth, 
  error bars/.cd, 
    y fixed,
    y dir=both, 
    y explicit
] table [x=x, y=y, col sep=comma] {data/basic/10/mushroom_3.txt};
\addlegendentry{$n=10$}

\addplot [name path=upper,draw=none,forget plot] table[x=x, y expr=\thisrow{y}+\thisrow{err},col sep=comma] {data/basic/10/mushroom_3.txt};
\addplot [name path=lower,draw=none,forget plot] table[x=x, y expr=\thisrow{y}-\thisrow{err},col sep=comma] {data/basic/10/mushroom_3.txt};
\addplot [fill=blue!10,fill opacity=0.8,forget plot] fill between[of=upper and lower];

\addplot+[
  green, mark options={scale=0.75},
  smooth, 
  error bars/.cd, 
    y fixed,
    y dir=both, 
    y explicit
] table [x=x, y=y, col sep=comma] {data/basic/100/mushroom_3.txt};
\addlegendentry{$n=100$}

\addplot [name path=upper,draw=none,forget plot] table[x=x, y expr=\thisrow{y}+\thisrow{err},col sep=comma] {data/basic/100/mushroom_3.txt};
\addplot [name path=lower,draw=none,forget plot] table[x=x, y expr=\thisrow{y}-\thisrow{err},col sep=comma] {data/basic/100/mushroom_3.txt};
\addplot [fill=green!10,fill opacity=0.8,forget plot] fill between[of=upper and lower];

\addplot+[
  brown, mark options={scale=0.75},
  smooth, 
  error bars/.cd, 
    y fixed,
    y dir=both, 
    y explicit
] table [x=x, y=y, col sep=comma] {data/basic/1000/mushroom_3.txt};
\addlegendentry{$n=1000$}

\addplot [name path=upper,draw=none,forget plot] table[x=x, y expr=\thisrow{y}+\thisrow{err},col sep=comma] {data/basic/1000/mushroom_3.txt};
\addplot [name path=lower,draw=none,forget plot] table[x=x, y expr=\thisrow{y}-\thisrow{err},col sep=comma] {data/basic/1000/mushroom_3.txt};
\addplot [fill=brown!10,fill opacity=0.8,forget plot] fill between[of=upper and lower];

\end{axis}
\end{tikzpicture}}
         \caption{Data 14.}
         \label{fig:d14}
     \end{subfigure}
     \hfill     
     \begin{subfigure}[b]{\figwidth}
         \centering
         \resizebox{\textwidth}{!}{\begin{tikzpicture}
\begin{axis}[
  grid=major,
  ymin=0, ymax=1, xmin=10, xmax=50,
  ytick align=outside, ytick pos=left,
  xtick align=outside, xtick pos=left,
  x label style={at={(axis description cs:0.5,-0.1)},anchor=north},
  y label style={at={(axis description cs:0,.5)},anchor=south},
  xlabel=Encoding size $\ell$,
  ylabel={Multiplicative Error $\epsilon$},
  legend pos=north east,
  legend style={draw=none}]
  
\addplot+[
  blue, mark options={scale=0.75},
  smooth, 
  error bars/.cd, 
    y fixed,
    y dir=both, 
    y explicit
] table [x=x, y=y, col sep=comma] {data/basic/10/mushroom_4.txt};
\addlegendentry{$n=10$}

\addplot [name path=upper,draw=none,forget plot] table[x=x, y expr=\thisrow{y}+\thisrow{err},col sep=comma] {data/basic/10/mushroom_4.txt};
\addplot [name path=lower,draw=none,forget plot] table[x=x, y expr=\thisrow{y}-\thisrow{err},col sep=comma] {data/basic/10/mushroom_4.txt};
\addplot [fill=blue!10,fill opacity=0.8,forget plot] fill between[of=upper and lower];

\addplot+[
  green, mark options={scale=0.75},
  smooth, 
  error bars/.cd, 
    y fixed,
    y dir=both, 
    y explicit
] table [x=x, y=y, col sep=comma] {data/basic/100/mushroom_4.txt};
\addlegendentry{$n=100$}

\addplot [name path=upper,draw=none,forget plot] table[x=x, y expr=\thisrow{y}+\thisrow{err},col sep=comma] {data/basic/100/mushroom_4.txt};
\addplot [name path=lower,draw=none,forget plot] table[x=x, y expr=\thisrow{y}-\thisrow{err},col sep=comma] {data/basic/100/mushroom_4.txt};
\addplot [fill=green!10,fill opacity=0.8,forget plot] fill between[of=upper and lower];

\addplot+[
  brown, mark options={scale=0.75},
  smooth, 
  error bars/.cd, 
    y fixed,
    y dir=both, 
    y explicit
] table [x=x, y=y, col sep=comma] {data/basic/1000/mushroom_4.txt};
\addlegendentry{$n=1000$}

\addplot [name path=upper,draw=none,forget plot] table[x=x, y expr=\thisrow{y}+\thisrow{err},col sep=comma] {data/basic/1000/mushroom_4.txt};
\addplot [name path=lower,draw=none,forget plot] table[x=x, y expr=\thisrow{y}-\thisrow{err},col sep=comma] {data/basic/1000/mushroom_4.txt};
\addplot [fill=brown!10,fill opacity=0.8,forget plot] fill between[of=upper and lower];

\end{axis}
\end{tikzpicture}}
         \caption{Data 15.}
         \label{fig:d15}
     \end{subfigure}
     \hfill     
     \begin{subfigure}[b]{\figwidth}
         \centering
         \resizebox{\textwidth}{!}{\begin{tikzpicture}
\begin{axis}[
  grid=major,
  ymin=0, ymax=1, xmin=10, xmax=50,
  ytick align=outside, ytick pos=left,
  xtick align=outside, xtick pos=left,
  x label style={at={(axis description cs:0.5,-0.1)},anchor=north},
  y label style={at={(axis description cs:0,.5)},anchor=south},
  xlabel=Encoding size $\ell$,
  ylabel={Multiplicative Error $\epsilon$},
  legend pos=north east,
  legend style={draw=none}]
  
\addplot+[
  blue, mark options={scale=0.75},
  smooth, 
  error bars/.cd, 
    y fixed,
    y dir=both, 
    y explicit
] table [x=x, y=y, col sep=comma] {data/basic/10/lymphography.txt};
\addlegendentry{$n=10$}

\addplot [name path=upper,draw=none,forget plot] table[x=x, y expr=\thisrow{y}+\thisrow{err},col sep=comma] {data/basic/10/lymphography.txt};
\addplot [name path=lower,draw=none,forget plot] table[x=x, y expr=\thisrow{y}-\thisrow{err},col sep=comma] {data/basic/10/lymphography.txt};
\addplot [fill=blue!10,fill opacity=0.8,forget plot] fill between[of=upper and lower];

\addplot+[
  green, mark options={scale=0.75},
  smooth, 
  error bars/.cd, 
    y fixed,
    y dir=both, 
    y explicit
] table [x=x, y=y, col sep=comma] {data/basic/100/lymphography.txt};
\addlegendentry{$n=100$}

\addplot [name path=upper,draw=none,forget plot] table[x=x, y expr=\thisrow{y}+\thisrow{err},col sep=comma] {data/basic/100/lymphography.txt};
\addplot [name path=lower,draw=none,forget plot] table[x=x, y expr=\thisrow{y}-\thisrow{err},col sep=comma] {data/basic/100/lymphography.txt};
\addplot [fill=green!10,fill opacity=0.8,forget plot] fill between[of=upper and lower];

\addplot+[
  brown, mark options={scale=0.75},
  smooth, 
  error bars/.cd, 
    y fixed,
    y dir=both, 
    y explicit
] table [x=x, y=y, col sep=comma] {data/basic/1000/lymphography.txt};
\addlegendentry{$n=1000$}

\addplot [name path=upper,draw=none,forget plot] table[x=x, y expr=\thisrow{y}+\thisrow{err},col sep=comma] {data/basic/1000/lymphography.txt};
\addplot [name path=lower,draw=none,forget plot] table[x=x, y expr=\thisrow{y}-\thisrow{err},col sep=comma] {data/basic/1000/lymphography.txt};
\addplot [fill=brown!10,fill opacity=0.8,forget plot] fill between[of=upper and lower];

\end{axis}
\end{tikzpicture}}
         \caption{Data 16.}
         \label{fig:d16}
     \end{subfigure}
        \caption{Multiplicative error of \fedchi w.r.t.~size of encoding $\ell$.}
        \label{fig:main}
\vspace{-20pt}
\end{figure}

\newcommand{\figwidthb}{0.3\textwidth}%
\tikzset{font={\fontsize{15pt}{12}\selectfont}}
\captionsetup{font={footnotesize},skip=2pt}
\captionsetup[sub]{font={footnotesize},skip=2pt}
\begin{wrapfigure}{r}{\figwidthb}
    \vspace{-15pt}
    \centering
    \resizebox{\figwidthb}{!}{\begin{tikzpicture}
\begin{axis}[
  grid=major,
  xtick distance=100, xticklabel style = {font=\Large},
  ytick distance=10, yticklabel style = {font=\Large},
  ymin=0, ymax=40, xmin=10, xmax=500,
  ytick align=outside, ytick pos=left,
  xtick align=outside, xtick pos=left,
  x label style={at={(axis description cs:0.5,-0.1)},anchor=north,font=\LARGE},
  y label style={at={(axis description cs:0,.5)},anchor=south,font=\LARGE},
  xlabel= {$m_x$ ($m_y$)},
  ylabel={Average Encoding Time (ms)},
  legend pos=north east,
  legend style={draw=none}]

\addplot+[
  blue, mark options={scale=0.25},
  smooth, 
  error bars/.cd, 
    y fixed,
    y dir=both, 
    y explicit
] table [x=x, y expr=1000*\thisrow{y}, col sep=comma] {data/benchmark/mobile.txt};
\addplot [name path=upper,draw=none, forget plot] table[x=x, y expr=\thisrow{y} * 1000 +\thisrow{err} * 1000,col sep=comma] {data/benchmark/mobile.txt};
\addplot [name path=lower,draw=none, forget plot] table[x=x, y expr=\thisrow{y} * 1000 -\thisrow{err} * 1000,col sep=comma] {data/benchmark/mobile.txt};
\addplot [fill=blue!10] fill between[of=upper and lower];

\end{axis}
\end{tikzpicture}}
    \caption{Client-side encoding overhead when $m_x=m_y$.}
    \label{fig:mobile}
\vspace{-15pt}
\end{wrapfigure}
We first evaluate the accuracy drop incurred by the estimation error of \fedchi,
which is due to its stable projection scheme and geometric mean estimator
(see \S~\ref{sec:fedchi-proof}). \F~\ref{fig:main} reports the evaluation
results: each point is an average of ten independent runs and half of the
error bars are the standard variance of the ten runs. We interpret that the
accuracy drop is independent of the number of clients. On the other hand, we
observe that the larger the encoding size $\ell$, the smaller the multiplicative
error. This is intuitive: according to Theorem~\ref{thm:utility}, the
multiplicative error $\epsilon = \sqrt{\frac{c}{\ell} \log(1/\delta)}$ will
become smaller if $\ell$ gets larger. When $\ell=50$, the multiplicative
error $\epsilon\approx0.2$. As the $\chi^2$ values are typically far from the
decision threshold, a multiplicative error of $0.2$ rarely flips the final
decision. This is further demonstrated in \S~\ref{sec:case}.

\vspace{-0.6em}
\paragraph{Client-side Computation Overhead.}
\tikzset{font={\fontsize{15pt}{12}\selectfont}}
\captionsetup{font={footnotesize},skip=2pt}
\captionsetup[sub]{font={footnotesize},skip=2pt}
\begin{wrapfigure}{r}{\figwidthb}
    \vspace{-10pt}
    \centering
    \resizebox{\figwidthb}{!}{\begin{tikzpicture}
\begin{axis}[
  grid=major,
  xtick distance=30,
  ymin=40, ymax=80, xmin=0, xmax=150,
  ytick align=outside, ytick pos=left, yticklabel style = {font=\Large},
  xtick align=outside, xtick pos=left, xticklabel style = {font=\Large},
  x label style={at={(axis description cs:0.5,-0.1)},anchor=north,font=\LARGE},
  y label style={at={(axis description cs:0,.5)},anchor=south,font=\LARGE},
  xlabel= {Epoch},
  ylabel={Model Accuracy (\%)},
  legend pos=south east,
  legend style={draw=none, font=\Large}]

\addplot+[
  brown, mark options={scale=0.1},
  smooth, 
  error bars/.cd, 
    y fixed,
    y dir=both, 
    y explicit
] table [x=x, y expr=100*\thisrow{y}, col sep=comma] {data/feature_selection/reuters_orig.txt};
\addlegendentry{no feature selection}

\addplot+[
  red, mark options={scale=0.1},
  smooth, 
  error bars/.cd, 
    y fixed,
    y dir=both, 
    y explicit
] table [x=x, y expr=100*\thisrow{y}, col sep=comma] {data/feature_selection/reuters_fed.txt};
\addlegendentry{\fedchi}

\addplot+[
  blue, mark options={scale=0.1},
  smooth, 
  error bars/.cd, 
    y fixed,
    y dir=both, 
    y explicit
] table [x=x, y expr=100*\thisrow{y}, col sep=comma] {data/feature_selection/reuters_baseline.txt};
\addlegendentry{centralized $\chi^2$-test}

\end{axis}
\end{tikzpicture}}
     
    \caption{Accuracy of the model trained with features selected by \fedchi\ and centralized $\chi^2$-test.}
    \label{fig:featselect}
\vspace{-10pt}
\end{wrapfigure}
To assess extra computation overhead incurred by \fedchi\ at the client side, we
measure the execution time of the encoding scheme on an Android 10 mobile device
with a Snapdragon865 processor and 12GB RAM. We download the \fedchi\ client model
to the Android device via PyDroid~\cite{pydroid}.

The results are shown in \F~\ref{fig:mobile}. Each point is an average of
100 independent runs and corresponding error bars are also reported. The
overhead is generally negligible. For example, for a $500\times 500$ contingency
table, the encoding only takes less than 30ms. The overhead grows linearly
w.r.t.~$m_x$ ($m_y$) and thus grows quadratically in \F~\ref{fig:mobile}
where $m_x=m_y$.

\vspace{-0.8em}
\subsection{Downstream Use Case Study}
\label{sec:case}
\vspace{-0.6em}

\paragraph{Feature Selection.}
%

Our first case study explores privacy-preserving feature selection on the basis
of \fedchi. In our setting, each client holds data with a large feature space.
The clients need to collaborate and run \fedchi\ to rule out unimportant
features and retain the top-$k$ features with the highest $\chi^2$ scores.

The dataset we use is the Reuters-21578~\cite{hayes1990construe}, which is a
standard dataset for text categorization~\cite{yang1999evaluation,
  yang1997comparative, zhang2003robustness}. We select the top 20 most frequent
categories with 17,262 training and 4,316 test documents. These documents are
randomly distributed to 100 clients, where each client holds the same number of
training documents. After ruling out all numbers and stop-words, we obtain
167,135 indexing terms. After conducting feature selection with \fedchi, we
select the top 40,000 terms with the highest $\chi^2$ scores. Compared with the
results of conducting centralized $\chi^2$-test, 38,012 (95.03\%) of the
selected terms are identical, illustrating that \fedchi has highly consistent
results with the standard $\chi^2$-test.

\tikzset{font={\fontsize{15pt}{12}\selectfont}}
\captionsetup{font={footnotesize},skip=2pt}
\captionsetup[sub]{font={footnotesize},skip=2pt}
\begin{wrapfigure}{r}{\figwidthb}
    \vspace{-10pt}
    \centering
    \resizebox{\figwidthb}{!}{\begin{tikzpicture}
\begin{axis}[
  grid=major,
  ymin=40, ymax=105, xmin=10, xmax=50,
  ytick align=outside, ytick pos=left, yticklabel style = {font=\Large},
  xtick align=outside, xtick pos=left, xticklabel style = {font=\Large},
  x label style={at={(axis description cs:0.5,-0.1)},anchor=north,font=\LARGE},
  y label style={at={(axis description cs:0,.5)},anchor=south,font=\LARGE},
  xlabel= {Encoding Size $\ell$},
  ylabel={Correct Rate (\%)},
  legend pos=south east,
  legend style={draw=none, font=\Large}]

\addplot+[
  blue, mark options={scale=0.75},
  smooth, line width=2pt,
  error bars/.cd, 
    y fixed,
    y dir=both, 
    y explicit
] table [x=x, y=y, col sep=comma] {data/cryptanalysis/1000.txt};
\addlegendentry{$L=1000$}

\addplot+[
  red, mark options={scale=0.75},
  smooth, dashed, line width=2pt, 
  error bars/.cd, 
    y fixed,
    y dir=both, 
    y explicit
] table [x=x, y=y, col sep=comma] {data/cryptanalysis/10000.txt};
\addlegendentry{$L=10000$}

\addplot+[
  green, mark options={scale=0.75},
  smooth, dashed, line width=2pt,
  error bars/.cd, 
    y fixed,
    y dir=both, 
    y explicit
] table [x=x, y=y, col sep=comma] {data/cryptanalysis/100000.txt};
\addlegendentry{$L=100000$}

\end{axis}
\end{tikzpicture}}
     
    \caption{The success rate of cracking Caesar ciphers. 
    }
    \label{fig:cryptanalysis}
\vspace{-25pt}
\end{wrapfigure}
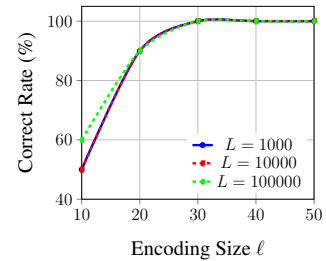
We then train logistic regression models with the terms selected by \fedchi\ and
centralized $\chi^2$-test, respectively. All other parameters like learning
rates and initial states are the same. The details of these models are reported
in Appendix~\ref{sec:model-detail}.
The results shown in \F~\ref{fig:featselect} further demonstrate that
\fedchi\ exhibits highly promising performance in feature selection which is
comparable to the centralized $\chi^2$-test.
We also evaluate the performance without feature selection, and as expected,
model accuracy after feature selection is higher than that without feature
selection. Note that the model without feature selection has 2,542,700 more
parameters than the model with feature selection; model after feature selection
can thus become much less complicated and save computing resources.

\vspace{-0.6em}
\paragraph{Cryptanalysis.}
In the second case study, we explore federated cryptanalysis with \fedchi. We
use Caesar cipher~\cite{caesar}, a classic substitution cipher where each letter
in the plaintext is replaced by another letter with some fixed number of positions
down the alphabet. For instance, each English letter can be right shift by 3,
thus mapping a plaintext ``good'' into a ciphertext ``jrrg''. Given in total 26
English letters, there are 26 possible shifts (the 26th ``shift'' maps a letter
to itself). The plaintext can be cracked in a shortcut by performing
correlation test on the ciphertext w.r.t.~normal English text.


In our setting, we assume each client holds a segment of the Caesar ciphertext.
To crack the ciphertext collaboratively, these clients run 26 $\chi^2$-tests to
compute the correlation level between each ciphertext letter and
letters in normal English text. The $\chi^2$-test yielding the highest
correlation level reveals how English letters are encrypted into Caesar
ciphertexts.

We pick Shakespeare's lines as the plaintext and encrypt it into Caesar
ciphertext. Let $L$ be the length of the ciphertext. We launch the cracking
process on ten non-overlapping ciphertexts to compute the average success rate
(see \F~\ref{fig:cryptanalysis}). In general, the longer ciphertext we have,
the easier to break it. A longer ciphertext can help better estimate the
frequency of each letter in ciphertext, thus boosting the correlation test. Also, the larger the encoding size,
the more accurate the $\chi^2$-statistics, and thus the higher the success rate
is. Again, according to Theorem~\ref{thm:utility}, the multiplicative error
$\epsilon$ decreases when the encoding size is larger.

\vspace{-0.6em}
\paragraph{Online False Discovery Rate Control.}

\tikzset{font={\fontsize{15pt}{12}\selectfont}}
\captionsetup{font={footnotesize},skip=2pt}
\captionsetup[sub]{font={footnotesize},skip=2pt}
\begin{wrapfigure}{r}{\figwidthb}
    \vspace{-10pt}
    \centering
    \resizebox{\figwidthb}{!}{\begin{tikzpicture}
\begin{axis}[
  grid=major,
  xtick distance=50,
  ymin=0, ymax=15, xmin=50, xmax=300,
  ytick align=outside, ytick pos=left, yticklabel style = {font=\Large},
  xtick align=outside, xtick pos=left, xticklabel style = {font=\Large},
  x label style={at={(axis description cs:0.5,-0.1)},anchor=north,font=\LARGE},
  y label style={at={(axis description cs:0,.5)},anchor=south,font=\LARGE},
  xlabel= {Encoding Size $\ell$},
  ylabel={FDR (\%)},
  legend pos=north east,
  legend style={draw=none, font=\Large}]

\addplot+[
  blue, mark options={scale=0.75},
  smooth, 
  error bars/.cd, 
    y fixed,
    y dir=both, 
    y explicit
] table [x=x, y expr=100*\thisrow{y}, col sep=comma] {data/fdr/fdr_gaussian.txt};
\addplot [name path=upper,draw=none, forget plot] table[x=x, y expr=\thisrow{y} * 100 +\thisrow{err} * 100,col sep=comma] {data/fdr/fdr_gaussian.txt};
\addplot [name path=lower,draw=none, forget plot] table[x=x, y expr=\thisrow{y} * 100 -\thisrow{err} * 100,col sep=comma] {data/fdr/fdr_gaussian.txt};
\addplot [fill=blue!10] fill between[of=upper and lower];

\end{axis}
\end{tikzpicture}}

    \caption{Average FDR w.r.t. encoding size $\ell$ for SAFFRON with \fedchi.}
    \label{fig:fdr}
\vspace{-10pt}
\end{wrapfigure}
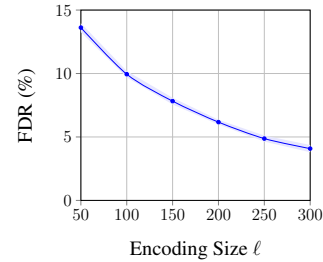

In the third case study, we explore federated online false discovery rate (FDR)
control \cite{foster2008alpha} with \fedchi.
In an online FDR control problem, a data analyst receives a stream of hypotheses on
the database, or equivalently a stream of $p$-values: $p_1, p_2, \cdots$.
At each time $t$, the data analyst should pick a threshold $\alpha_t$ to reject
the hypothesis when $p_t < \alpha_t$.
The error metric is the false discovery rate, and the goal of online FDR control
is to guarantee that for any time $t$, the FDR up to time $t$ is less than a
pre-determined quantity.

We use the SAFFRON procedure~\cite{ramdas2018saffron}, the state-of-the-art
online FDR control for multiple hypothesis testing. The $\chi^2$ results and
corresponding $p$-values are calculated by \fedchi. We present the detailed
algorithm of SAFFRON and the hyper-parameters used in our evaluation in Appendix~\ref{sec:SAFF}.
The size of the randomly synthesized contingency table is $20\times 20$. There
are $100$ independent hypotheses each time and the probability of each
hypothesis to be independent or correlated is $0.5$. The time sequence length is
$100$ and the number of clients is $10$.
%
The data is synthesized from multivariate Gaussian distribution. For the
correlated data, the covariance matrix is randomly sampled from a uniform
distribution, whereas for the independent data, the covariance matrix is
diagonal whose entries are randomly sampled from a uniform distribution.

At time $t$, we use \fedchi\ to calculate the $p$-values $p_t$ of all the
hypotheses, and then SAFFRON procedure takes $p_t$ to estimate the reject
threshold $\alpha_t$.
The relationship between the average FDR and encoding size $\ell$ is shown in
\F~\ref{fig:fdr}, where each point is the average of five independent runs
with different random seeds and half of the error bar is the standard variance
of the five runs.
Low FDR, which is less than $5.0\%$, can be achieved by increasing the encoding
size $\ell$. This further demonstrates that \fedchi\ can be employed in practice
to facilitate online FDR control with privacy-preserving correlation test.

\vspace{-0.8em}
\section{Limitation \& Conclusion}
\label{sec:conclusion}
\vspace{-0.6em}

This paper takes an important step towards designing non-linear secure
aggregation protocols in the federated setting. Specifically, we propose a
universal secure protocol to evaluate frequency moments in the federated
setting. We focus on an important application of the protocol: $\chi^2$-test. We
give formal security proof and utility analysis on our proposed secure federated
learning $\chi^2$-test protocol \fedchi\ and validate them with empirical
evaluation.

We now discuss limitations and potential future works of \fedchi. Other correlation tests like G-test can also be decomposed into frequency moments
estimation. However, the precision of the federated G-test protocol can be
likely low, due to the structural bias in the approximation and computational
imprecision.
We deem it as an important future direction to deliver privacy-preserving
solutions for other correlation tests. Second, if the dimensions of the
contingency table are too small to hide the joint distribution, the security
guarantee of \fedchi will be undermined and might leave a chance for attackers
to perform inference attack. Our observation shows that $m = 400, m_x = m_y =
20, \ell = 50$ is a good configuration with promising accuracy. In practice, we
require that $m > m_x + m_y + \ell$ and the optimal choice should generally
depend on the requirement of the security level, as well as the specific task.

\bibliographystyle{plain}
\bibliography{ref.bib}

\appendix
\section*{Appendix}
\section{Secure Aggregation}
\label{sec:secureagg}

The secure aggregation protocol from~\cite{bell2020secure} is presented in \A~\ref{alg:secureagg}. 
The first step of the protocol is to generate a $k$-regular graph $G$, where the $n$ vertices are the clients participating in the protocol.
The server runs a randomized graph generation algorithm {\sc InitSecureAgg} presented in \A~\ref{alg:secureagginit} that takes the number of clients $n$ and samples output $(G, k)$ from a distribution $\mathcal{D}$. In \A~\ref{alg:secureagginit}, we uniformly rename the nodes of a graph known as a Harary graph defined in Definition~\ref{def:harary} with $n$ nodes and $k$ degrees. The graph $G$ is constructed by sampling $k$ neighbours uniformly and without replacement from the set of remaining $n-1$ clients. We choose $k = \mathcal{O}(log(n))$, which is large enough to hide the updates inside the masks.

In the second step, the edges of the graph determine pairs of clients, each of which runs a key agreement protocol to share a random key. The random key will be used by each party to derive a mask for her input.

In the third step, each pair $(i,j)$ of connected clients in $G$ runs a $\lambda$-secure key agreement protocol $s_{i,j} = \mathcal{KA}.Agree(sk_i^1, pk_j^1)$ which uses the key exchange in the previous step to derive a shared random key $s_{i, j}$.
The pairwise masks $\textbf{m}_{i,j} = F(s_{i,j})$ can be computed, where $F$ is the pseudorandom generator (PRG).
Further, the masks are added to the input of client $c_i$:  $\textbf{y}_i = \textbf{e}_i - \sum_{j \in [n], j < i} \textbf{m}_{i,j} + \sum_{j \in [n], j > i} \textbf{m}_{i,j}$.

Finally, the server can cancel out the pairwise masks in the final sum: $\sum_{i \in [n]} \textbf{y}_i$.

\begin{definition}[\textsc{Harary}$(n, k)$ Graph]
\label{def:harary}
Let \textsc{Harary}$(n, k)$ denotes a graph with $n$ nodes and degree $k$. This graph has vertices $V = [n]$ and an edge between two distinct vertices $i$ and $j$ if and only if $j - i \pmod{n} \leq (k+1)/2$ or $j - i \pmod{n} \geq n - k/2$.
\end{definition}

\begin{algorithm}
  \caption{\textsc{InitSecureAgg}: Generate Initial Graph for \textsc{SecureAgg}.}
  \label{alg:secureagginit}
  \DontPrintSemicolon
  \SetKwProg{Fn}{Function}{:}{}
  \SetKwFunction{FS}{\textsc{SecureAgg}}
  \SetKwFunction{FI}{\textsc{InitSecureAgg}}
  \SetKwFunction{CD}{\textsc{ComputeDegreeAndThreshold}}
  \SetKwFunction{HA}{\textsc{Harary}}
  \Fn{\FI{$n$}}{
        $\triangleright$ $n$: Number of nodes.\;
        $k = \mathcal{O}(log(n))$.\;
        Let $H =$ \HA{$n, k$}.\;
        Sample a random permutation $\pi : [n] \rightarrow [n]$.\;
        Let $G$ be the set of edges $\{ (\pi(i), \pi(j)) | (i, j) \in H \}$.\;
        \KwRet $(G, k)$
  }
\end{algorithm}

\begin{algorithm}
  \caption{\textsc{SecureAgg}: Secure Aggregation Protocol.}
  \label{alg:secureagg}
  \DontPrintSemicolon
  \SetKwProg{Fn}{Function}{:}{}
  \SetKwFunction{FS}{\textsc{SecureAgg}}
  \SetKwFunction{FI}{\textsc{InitSecureAgg}}
  \Fn{\FS{$\{\textbf{e}_i\}_{i \in [n]}$}}{
        $\triangleright$ Parties: Clients $c_1, \cdots, c_n$, and Server.\;
        $\triangleright$ $l$: Vector length.\;
        $\triangleright$ $\mathbb{X}^l$: Input domain, $\textbf{e}_i \in \mathbb{X}^l$.\;
        $\triangleright$ $F: \{0, 1\}^{\lambda} \rightarrow \mathbb{X}^l$: PRG.\;
        (1) The server runs $(G, k) = \FI{n}$, where $G$ is a regular degree-$k$ undirected graph with $n$ nodes. By $N_G(i)$ we denote the set of $k$ nodes adjacent to $c_i$ (its neighbors).\;
        (2) Client $c_i, i \in [n]$, generates key pairs $(sk_i^1, pk_i^1)$ and sends $pk_i^1$ to the server who forwards the message to $N_G(i)$.\;
        
        (3) \For{each Client $c_i, i \in [n]$}{
            \noindent {\tikz\draw[black,fill=black] (-0.5em,-0.5em) circle (-0.15em);} Computes a shared random PRG seed $s_{i,j}$ as $s_{i,j} = \mathcal{KA}.Agree(sk_i^1, pk_j^1)$.\;
            \noindent {\tikz\draw[black,fill=black] (-0.5em,-0.5em) circle (-0.15em);} Computes masks $\textbf{m}_{i,j} = F(s_{i,j})$.\;
            \noindent {\tikz\draw[black,fill=black] (-0.5em,-0.5em) circle (-0.15em);} Sends to the server their masked input
            $$ \textbf{y}_i = \textbf{e}_i - \sum_{j \in [n], j < i} \textbf{m}_{i,j} + \sum_{j \in [n], j > i} \textbf{m}_{i,j} $$
        }
        
        (4) The server collects masked inputs and 
        \KwRet $ \sum_{i \in [n]} \textbf{y}_i$
  }
\end{algorithm}
\section{Proof for Communication \& Computation Cost}
\label{sec:proof-cost}

We provide the proof for Theorem~\ref{thm:comm} and Theorem~\ref{thm:comp} in the following.

\textbf{Theorem}~\ref{thm:comm} (Communication Cost).
Let $\Pi$ be an instantiation of \A~\ref{alg:main} with secure aggregation protocol from~\cite{bell2020secure}, then (1) the client-side communication cost is $\mathcal{O}(\log n + m_x + m_y + \ell)$; (2) the server-side communication cost $\mathcal{O}(n\log n + nm_x + nm_y + n\ell)$.

\begin{proof}
Each client performs $k$ key agreements ($\mathcal{O}(k)$ messages, line 9 in \A~\ref{alg:secureagg}) and sends 3 masked inputs ($\mathcal{O}(m_x + m_y + \ell)$ complexity, line 3, 4, 15 in \A~\ref{alg:main} and line 10 in \A~\ref{alg:secureagg}). Thus the client communication cost is $\mathcal{O}(\log n + m_x + m_y + \ell)$.

The server receives or sends $\mathcal{O}(\log n + m_x + m_y + \ell)$ messages to each client, so the server communication cost is $\mathcal{O}(n\log n + nm_x + nm_y + n\ell)$.
\end{proof}

\textbf{Theorem}~\ref{thm:comp} (Computation Cost).
Let $\Pi$ be an instantiation of \A~\ref{alg:main} with secure aggregation protocol from~\cite{bell2020secure}, then (1) the client-side computation cost is 
$\mathcal{O}(m_x\log n + m_y \log n + \ell\log n + m\ell)$; 
%
(2) the server-side computation cost is 
$\mathcal{O}(m_x + m_y + \ell)$.

\begin{proof}
Each client computation can be broken up as $k$ key agreements ($\mathcal{O}(k)$ complexity, line 9 in \A~\ref{alg:secureagg}), generating masks $\textbf{m}_{i,j}$ for all neighbors $c_j$ ($\mathcal{O}(k(m_x + m_y + \ell))$ complexity, line 3, 4, 15 in \A~\ref{alg:main} and line 10 in \A~\ref{alg:secureagg}), and encoding computation cost $\mathcal{O}(m\ell)$ (line 14 in \A~\ref{alg:main}). Thus the client computation cost is $\mathcal{O}(m_x\log n + m_y \log n + \ell\log n + m\ell)$.

The server computes 3 aggregated results by summing the received masked results, thus the server computation cost is $\mathcal{O}(m_x + m_y + \ell)$ (line 3, 4, 15 in \A~\ref{alg:main} and line 12 in \A~\ref{alg:secureagg}).
\end{proof}
\section{Details of Datasets}
\label{sec:dataset-detail}
The details for the real-world datasets used in \S~\ref{subsec:eval-results} are provided in Table~\ref{tab:data}.
The license of Credit Risk Classification~\cite{credit} is CC BY-SA 4.0, the license of German Traffic Sign~\cite{Houben-IJCNN-2013} is CC0: Public Domain. Other datasets without a license are from UCI Machine Learning Repository~\cite{Dua:2019}.

\begin{table}[
htbp]
    \caption{Dataset details.}
    \label{tab:data}
    \centering
    \small
    \resizebox{\columnwidth}{!}{
    \begin{tabular}{|c|c|c|c|c|c|}
    \hline
    ID & Data & Attr \#1 & A\#1 Cat & Attr \#2 & A\#2 Cat \\
    \hline
    1 & Adult Income~\cite{kohavi1996scaling,adult} & Occupation & 14 & Native Country & 41\\
    \hline
    2 & Credit Risk Classification~\cite{credit} & Feature 6 & 14 & Feature 7 & 11 \\
    \hline
    3 & Credit Risk Classification~\cite{credit} & Credit Product Type & 28 & Overdue Type I & 35 \\
    \hline
    4 & Credit Risk Classification~\cite{credit} & Credit Product Type & 28 & Overdue Type II & 35 \\
    \hline
    5 & Credit Risk Classification~\cite{credit} & Credit Product Type & 28 & Overdue Type III & 36 \\
    \hline
    6 & German Traffic Sign~\cite{Houben-IJCNN-2013} & Image Width & 219 & Traffic Sign & 43\\
    \hline
    7 & German Traffic Sign~\cite{Houben-IJCNN-2013} & Image Height & 201 & Traffic Sign & 43\\
    \hline
    8 & German Traffic Sign~\cite{Houben-IJCNN-2013} & Upper left X coordinate & 21 & Traffic Sign & 43\\
    \hline
    9 & German Traffic Sign~\cite{Houben-IJCNN-2013} & Upper left Y coordinate & 16 & Traffic Sign & 43\\
    \hline
    10 & German Traffic Sign~\cite{Houben-IJCNN-2013} & Lower right X coordinate & 204 & Traffic Sign & 43\\
    \hline
    11 & German Traffic Sign~\cite{Houben-IJCNN-2013} & Lower right Y coordinate & 186 & Traffic Sign & 43\\
    \hline
    12 & Mushroom~\cite{mushroom} & Cap color & 10 & Odor & 9 \\
    \hline
    13 & Mushroom~\cite{mushroom} & Gill color & 12 & Stalk color above ring & 9 \\
    \hline
    14 & Mushroom~\cite{mushroom} & Stalk color below ring & 9 & Ring Type & 8 \\
    \hline
    15 & Mushroom~\cite{mushroom} & Spore print color & 9 & Habitat & 7\\
    \hline
    16 & Lymphography~\cite{lymphography} & Structure Change & 8 & No. of nodes & 8 \\
    \hline
    \end{tabular}
    }
\end{table}
\section{Details of Regression Models}
\label{sec:model-detail}

The details of the regression models trained in feature selection in \S~\ref{sec:case} is reported in Table~\ref{tab:model}. 
The training and testing splits are the same for \fedchi, centralized $\chi^2$-test and model without feature selection (i.e. there are 17,262 training and 4,316 test documents). 
We use the same learning rate; random seed and all other settings are also the same to make the comparison fair. We get the result of \F~\ref{fig:featselect} and the models are all trained on NVIDIA GeForce RTX 3090.

\begin{table}[
htbp]
    \caption{Model details.}
    \label{tab:model}
    \centering
    \small
    \begin{tabular}{|c|c|c|c|c|c|}
    \hline
    Task & Model Size & Learning Rate & Random Seed \\
    \hline
    \fedchi\ & $40000 \times 20$ & $0.1$ & $0$ \\
    \hline
    Centralized $\chi^2$-test & $40000 \times 20$ & $0.1$ & $0$ \\
    \hline
    Without Feature Selection & $167135 \times 20$ & $0.1$ & $0$ \\
    \hline
    \end{tabular}
\end{table}

\section{SAFFRON Procedure}
\label{sec:SAFF}

In \S~\ref{sec:case}, we adopt SAFFRON procedure~\cite{ramdas2018saffron} to perform online FDR control. 
SAFFRON procedure is currently state-of-the-art for multiple hypothesis testing. 
In \A~\ref{alg:saff}, we formally present the SAFFRON algorithm. 

\begin{algorithm}
  \caption{SAFFRON Procedure.}
  \label{alg:saff}
  \DontPrintSemicolon
  \SetKwFunction{FS}{\textsc{SaffronProcedure}}
  \SetKwProg{Fn}{Function}{:}{}

  \Fn{\FS{$\{p_1, p_2, \cdots\}$, $\alpha$, $W_0$, $\{\gamma_j\}_{j=0}^\infty$}}{
        $\triangleright$ $\{p_1, p_2, \cdots\}$: Stream of $p$-values.\;
        $\triangleright$ $\alpha$: Target FDR level.\;
        $\triangleright$ $W_0$: Initial wealth.\;
        $\triangleright$ $\{\gamma_j\}_{j=0}^\infty$: Positive non-increasing sequence summing to one.\;
        $i \leftarrow 0$ \tcp*{Set rejection number.}
        \For{each $p$-value $p_t \in \{p_1, p_2, \cdots\}$}{
            $\lambda_t \leftarrow g_t(R_{1:t-1}, C_{1:t-1})$\;
            $C_t \leftarrow I(p_t < \lambda_t)$ \tcp*{Set the indicator for candidacy $C_t$.}
            $C_{j+} \leftarrow \sum_{i=\tau_j+1}^{t-1}C_i$ \tcp*{Set the candidates after the $j^{th}$ rejection.}
            \If{$t = 1$}{
                $\alpha_1 \leftarrow (1 - \lambda_1) \gamma_1 W_0$\;
            }
            \Else{
                $\alpha_t \leftarrow (1 - \lambda_t) (W_0 \gamma_{t - C_{0+}} + (\alpha - W_0) \gamma_{t - \tau_1 - C_{1+}} + \sum_{j \geq 2} \alpha \gamma_{t - \tau_j - C_{j+}})$\;
            }
            $R_t \leftarrow I(p_t \leq \alpha_t)$ \tcp*{Output $R_t$.}
            \If{$R_t = 1$}{
                $i \leftarrow i + 1$ \tcp*{Update rejection number.}
                $\tau_i \leftarrow t$ \tcp*{Set the $i^{th}$ rejection time.}
            }
        }
        \KwRet $\{R_0, R_1, \cdots\}$
  }
\end{algorithm}

The initial error budget for SAFFRON is $(1 - \lambda_1 W_0) < (1 - \lambda_1 \alpha)$, and this will be allocated to different tests over time. 
The sequence $\{\lambda_j\}_{j=1}^{\infty}$ is defined by $g_t$ and $\lambda_j$ serves as a weak estimation of $\alpha_j$.
$g_t$ can be any coordinate wise non-decreasing function (line 8 in \A~\ref{alg:saff}).
$R_j \coloneqq I(p_j < \alpha_j)$ is the indicator for rejection, while $C_j \coloneqq I(p_j < \lambda_j)$ is the indicator for candidacy.
$\tau_j$ is the $j^{th}$ rejection time. 
For each $p_t$, if $p_t < \lambda_t$, SAFFRON adds it to the candidate set $C_t$ and sets the candidates after the $j^{th}$ rejection (line 9-10 in \A~\ref{alg:saff}). 
Further, the $\alpha_t$ is updated by several parameters like current wealth, current total rejection numbers, the current size of the candidate set, and so on (line 11-14 in \A~\ref{alg:saff}).
Then the decision $R_t$ is made according to the updated $\alpha_t$ (line 15 in \A~\ref{alg:saff}).

The hyper-parameters we use for SAFFRON procedure in online false discovery rate control of \S~\ref{sec:eval} are aligned with the setting in~\cite{ramdas2018saffron}, and they are listed below: the target FDR level is $\alpha=0.05$, the initial wealth is $W_0 = 0.0125$ and $\gamma_j$ is calculated in the following way: $\gamma_j = \frac{1 / (j + 1)^{1.6}}{\sum_{j=0}^{10000} 1/(j + 1)^{1.6}}$.

\end{document}